\documentclass[sigconf]{acmart}

\AtBeginDocument{%
  \providecommand\BibTeX{{%
    \normalfont B\kern-0.5em{\scshape i\kern-0.25em b}\kern-0.8em\TeX}}}

\usepackage{stmaryrd,dsfont,amsmath,amsfonts}
\usepackage{semantic,multirow,multicol,graphicx}
\usepackage[algoruled,linesnumbered,noend]{algorithm2e}
\usepackage{algorithmic}
\usepackage{graphicx}
\usepackage{tikz}
\usepackage{pifont}
\usepackage{threeparttable}
\usepackage{diagbox}
\usepackage{xspace}
\usepackage{amsmath}
\usepackage{enumitem}
\usepackage{url,comment}
\usepackage{subfigure}
\usepackage{tcolorbox}

\usetikzlibrary{automata,arrows,positioning}

\newcommand{\tool}{{A$^2$D}\xspace}%

\newcommand{\fu}[1]{\textcolor{blue}{#1}}

\newcommand{\zhe}[1]{\textcolor{red} {ZHE: #1}}

\newcommand{\tabincell}[2]{
\begin{tabular}{@{}#1@{}}#2\end{tabular}
}

\begin{document}

\title{Attack as Defense: Characterizing Adversarial Examples using Robustness}
 
\author{Zhe Zhao}
\affiliation{\institution{ShanghaiTech University}\country{China} }

\author{Guangke Chen}
\affiliation{\institution{ShanghaiTech University}\country{China}} 

\author{Jingyi Wang} 
\affiliation{\institution{Zhejiang University}\country{China} }

\author{Yiwei Yang}
\affiliation{\institution{ShanghaiTech University}\country{China}}

\author{Fu Song}
\affiliation{\institution{ShanghaiTech University}\country{China}}

\author{Jun Sun}
\affiliation{\institution{Singapore Management University}\country{Singapore} } 

\begin{abstract}
As a new programming paradigm, deep learning has expanded its application to many real-world problems. At the same time, deep learning based software are found to be vulnerable to adversarial attacks. Though various defense mechanisms have been proposed to improve robustness of deep learning software, many of them are ineffective against adaptive attacks.
In this work, we propose a novel characterization to distinguish adversarial examples from benign
ones based on the observation that adversarial examples are significantly less robust than benign ones.
As existing robustness measurement does not
scale to large networks,
we propose a novel defense framework, named attack as defense (\tool), to detect adversarial examples by effectively evaluating an example's robustness.
\tool uses the cost of attacking an input for robustness evaluation and identifies those less robust examples as adversarial
since less robust examples are easier to attack.
Extensive experiment results on MNIST, CIFAR10 and ImageNet show that \tool is more effective than recent promising approaches.
We also evaluate our defence against potential adaptive attacks and
show that \tool is effective in defending
carefully designed adaptive attacks, e.g., the attack success rate drops to 0\% on CIFAR10.
\end{abstract}

\maketitle

\section{Introduction}
\label{sec:intr}

Deep learning (DL) has arguably become a new programming paradigm which takes over traditional software programs in many areas.
For instance, it has
achieved state-of-the-art performance in real-world tasks
such as autonomous driving~\cite{Apollo},
medical diagnostics~\cite{SWS17}
and cyber-security~\cite{SYLS18}.
Despite the success, DL software are still far from dependable (especially for safety- and security-critical systems) and, like traditional software, they must be properly tested, and defended in the presence of malicious inputs.
In particular, 
DL software are known to be brittle to adversarial examples~\cite{Intriguing,FGSM15,CW}, i.e., by adding a slight perturbation on an input, a well-trained DL model could be easily fooled.

There is a huge body of work proposing various attack and defense mechanisms with regard to adversarial examples, from both the security~\cite{FGSM15,KGB17,JSMA,BRB18} and software engineering community~\cite{wang2019adversarial,MJZSXLCSLLZW18,TPJR18,SWRHKK18,Wang2019Dissector}.
Since the adversarial attack L-BFGS was introduced~\cite{Intriguing}, many sophisticated adversarial attacks have been proposed, such as
Fast Gradient Sign Method (FGSM)~\cite{FGSM15}, Iterative Gradient Sign Method (BIM)~\cite{KGB17},
Jacobian-based Saliency Map Attack (JSMA)~\cite{JSMA}, Local Search Attack (LSA)~\cite{NK17}, Decision-Based Attack (DBA)~\cite{BRB18},
DeepFool~\cite{DeepFool}, and Carlini and Wagner's attack (C\&W)~\cite{CW}.
As countermeasures, defense mechanisms are proposed to improve robustness in the presence of adversarial attacks.
Examples include adversarial training~\cite{FGSM15,MyMSTV17,ShafahiNG0DSDTG19}, defensive distillation~\cite{PM0JS16}
and feature squeezing~\cite{Xu0Q18}. These works constitute important
steps in exploring the defense mechanisms, yet have various limitations and are shown to be insufficient~\cite{CW,HeWCCS17,CW17a,athalye2018obfuscated,TCBM20}.

Some efforts also have been made to distinguish adversarial examples from benign ones and reject those potentially adversarial ones~\cite{wang2019adversarial,Wang2019Dissector}. A usual underlying assumption is that adversarial and benign examples differ in a certain subspace distribution, e.g., kernel density~\cite{feinman2017detecting}, local intrinsic dimensionality~\cite{LID}, manifold~\cite{MC17}, logits values~\cite{RothKH19}, etc.
Although these characterizations provide some insights on the adversarial subspace, they are far from reliably discriminating adversarial examples alone.
Worse yet, attackers can easily break existing defense mechanisms by specifically designed adaptive attacks~\cite{CW17a,carlini2017magnet,TCBM20}.

In this work, we propose a novel characterization, i.e., robustness, to distinguish adversarial examples from benign ones.
Our main observation is that adversarial examples (crafted by most existing attack methods) are significantly less robust than benign ones.
Different from previous statistical characterizations~\cite{feinman2017detecting,LID,MC17,RothKH19},
the difference in robustness comes from two inherent characteristics of model training and adversarial example generation. First, the model training
progressively minimizes the loss of each example with respect to the model. Thus, a benign example is often trained to be robust against a model with good generalization. As a result, benign examples are in general relatively far away from the decision boundary. Second, most adversarial attacks aim to generate human-imperceptible perturbations which often result in much less robust just-cross-boundary adversarial examples. The contrasting robustness characteristics make it suitable to distinguish adversarial examples from benign ones.

However, it is still an open research problem on how to effectively measure an input's robustness with respect to a DL model. Existing robustness measurement methods (e.g., CLEVER \cite{Clever} which calculates a minimum perturbation required to change an input's label) are often too computationally expensive. In this work, we propose to utilize the robustness difference from a reversed angle. Our intuition is that it is easier to employ a successful adversarial attack on an input that is less robust. Thus, we propose a novel defense, named \emph{attack as defense} (\tool), to effectively detect adversarial examples from benign ones. Given an input example, we apply different kinds of attacks to attack the input and measure how `easy' it is to employ a successful attack. An input example is considered less robust and thus more likely to be adversarial, if the attacks are easier to succeed. Note that the effectiveness of our defense (\tool) relies on a set of attacks whose attack cost (easiness) can be quantitatively measured, which will guide the selection of attacks from a large number of adversarial attacks in the literature.

Compared to existing adversarial example detection approaches (which are mostly proven to be ineffective~\cite{athalye2018obfuscated,CW,CW17a,carlini2017magnet,HeWCCS17,TCBM20}),
our \tool framework has the following advantages. First, \tool utilizes the inherent robustness difference caused by the contrasting characteristics of model training and adversarial example generation.
To circumvent the defense, an attack needs to generate robust adversarial examples which in general might induce human-perceptible large distortion.
Second, \tool is essentially an ensemble approach using the robustness information obtained from different kinds of attacks which is hard to bypass at once.
We evaluated \tool with a carefully selected set of attacks on three popular datasets MNIST~\cite{MNIST98}, CIFAR10~\cite{Kri09} and ImageNet~\cite{TCBM20}.
Experiment results show that \tool can be more effective than recently proposed detection algorithms~\cite{feinman2017detecting,LID,wang2019adversarial,Wang2019Dissector} and remain effective even in the white-box adversarial setting.
We further show that \tool is effective against adaptive attacks. That is, we show that \tool (combined with a complementary detection approach for detecting large-distortion adversarial examples~\cite{MC17}
and adversarial training) is very effective against specifically designed adaptive attacks, e.g., the attack success rate (ASR) drops from
72\% to 0\% using our defense with adversarial training on CIFAR10, and the ASR drops from 100\% to 0\% on MNIST.
We remark that many existing defenses combined with adversarial training result in lower robustness than adversarial training on its own~\cite{TCBM20}.

In a nutshell, we make the following contributions:
\begin{itemize}
  \item We propose a novel characterization to distinguish adversarial examples from benign ones via robustness.
  \item We present detection approaches based on the characterization, which can utilize existing attacks and do not need to modify or retrain the protected model.
  \item We conduct extensive experiments to test our observations and our defense, which outperforms recent promising detection algorithms.
  \item We thoroughly discuss possible adaptive attacks to our defense and evaluate them to our defense integrated with a complementary detection approach and adversarial training. The integrated defense is very promising.
\end{itemize}


\section{Background}
\label{sec:back}


\subsection{Adversarial Attacks}
In this work, we target deep neural network (DNN) for classification tasks.
We denote a DNN by $f:X\to C$, which maps each input $x\in X$ to a certain label $c\in C$. We denote the ground-truth label of an input $x$ by $c_x$. Given a DNN $f$ and a benign input $x$ (which means $f(x)=c_x$),
the attacker's goal is to craft a perturbation $\Delta x$ (measured in different $L_n$ norms~\cite{CW}) for the input $x$ such that the DNN $f$ classifies the example $\Hat{x}=x+\Delta x$ as a different label $f(\Hat{x})$, i.e., $f(\Hat{x})\neq f(x)$. Such $\Hat{x}$ is called an adversarial example.

In the literature, an extensive number of adversarial attacks have been proposed~\cite{Intriguing,FGSM15,CW,JSMA,KGB17,EEF0RXPKS18,PMGJCS17,BRB18,DeepFool,DFA,TTCLZYC18,IEAL18,CSZYH18,NYC15,MFFF17}.
We briefly introduce some representative attacks that will be used for robustness evaluation in our work. 

\noindent \textbf{FGSM.}
Fast Gradient Sign Method (FGSM)~\cite{FGSM15} uses a loss function $J (x,c_x)$~(e.g. the cross-entropy loss)
describes the cost of classifying $x$ as label $c_x$,
and maximizes the loss to implement an untarget attack
by performing one step gradient ascend
from the input $x$ with a $L_\infty$ distance threshold $\epsilon$.
Formally, a potential adversarial example $\Hat{x}$ is crafted
as follows:
\begin{center}
$\Hat{x}=x+{\epsilon}\times\mathbf{sign}(\nabla_x{J(x,c_x)})$
\end{center}
where
$\nabla_x$ is the partial derivative of $x$,
and $\mathbf{sign}(\cdot)$ is a sign function such that $\mathbf{sign}(c)$ is $+1$ if $c>0$,
$-1$ if $c<0$ and $0$ if $c=0$.


\noindent \textbf{BIM.}
Basic Iterative gradient Method (BIM)~\cite{KGB17}
is an iterative version of FGSM.
For each iteration, BIM performs FGSM with a small step size $\alpha$
and clips the result so that it stays in the $\epsilon$-neighbourhood of the input sample.
The $i$th iteration is updated by as follows:
\begin{center}
$x^{i+1}={\tt clip}_{\epsilon,x}(x^i+\alpha\times\mathbf{sign}(\nabla_x{J(x^i,c_x)}))$
\end{center}
where $x^0=x$, and the iterative process can repeat several times.

The perturbation of FGSM and BIM is restricted by the $L_\infty$ norm,
measuring the largest change between $\hat{x}$ and $x$ (i.e. $\|x-\hat{x}\|_\infty \leq \epsilon$).
We could derive $L_2$ norm (i.e. $\|x-\hat{x}\|_2 \leq \epsilon$) attacks by,
\begin{center}
$\Hat{x}=x+{\epsilon}\times \frac{\nabla_x{J(x,c_x)}}{\nabla_x{\|J(x,c_x)\|_2}}$
\end{center}
Similarly, FGSM and BIM can be adapted from untarget attacks to target ones
which specify the target label of an adversarial example.

Compared to fixed step size,
there are some optimization-based attack methods  that
seek to find adversarial examples with the minimal perturbation,
such as {L-BFGS}~\cite{Intriguing} and {C\&W}~\cite{CW}.
In addition,
{JSMA}~\cite{JSMA} seeks to modify the smallest number of pixels,
which is an attack method with the $L_0$ norm.

\subsection{Robustness}
%
A DNN $f$ is (locally) \emph{robust} with respect to an input $x$ and an $L_p$ norm distance threshold $\epsilon$
if for every example $\hat{x}$ such that $\|x-\hat{x}\|_p\leq \epsilon$,  $f(x)=f(\hat{x})$ holds.
Several approaches have been proposed to certify robustness, based on SAT/SMT/MILP solving~\cite{Ehl17,KBDJK17}, abstraction refinement~\cite{wang2018efficient,WPWYJ18,EGK20},
and abstract interpretation~\cite{GMDTCV18,SGMPV18,SGPV19}.
Though these approaches feature theoretical guarantees,
they are limited in scalability and efficiency, hence fail to work for large models in practice.
To improve scalability, a few approaches aiming to achieve statistical guarantees, by claiming robustness with certain probability~\cite{Clever,BILVNC16,RHK18}.
Among them, CLEVER~\cite{Clever} score is an effective metric to estimate robustness
by sampling the norm of gradients and fitting a limit distribution
using extreme value theory. For each input $x$, CLEVER is able to calculate a minimal perturbation $\Delta x$ needed such that $\Hat{x}=x+\Delta x$ becomes an adversarial example of $x$.
The CLEVER score is an attack-independent robustness metric for large scale neural networks.
Readers can refer to~\cite{Clever} for details. In this work, we use the CLEVER score to compare the robustness of adversarial and benign examples.



\subsection{Problem Formulation}
We focus on the detection of adversarial examples as motivated by many relevant works~\cite{feinman2017detecting,LID,wang2019adversarial,Wang2019Dissector}.
%
The problem is: \textit{given an input example $x$ to a DNN model $f$, how to effectively decide whether $x$ is benign or adversarial?} The fundamental problem is how to better characterize adversarial examples. Our solution is to use robustness.


\smallskip
\noindent\textbf{Threat Model}. We consider a challenging defense scenario which assumes that the adversary knows all the information
of the model under attack, namely, white-box attacks.
Besides, we assume the detector has access to a set of benign examples, but knows nothing about how the adversary generates adversarial examples.
We also assume the detector can use various attacks (for robustness evaluation). These assumptions are reasonable in practice, as there are many publicly available datasets and source of attacks.



\section{Characterization}

\subsection{Robustness: Adversarial vs. Benign}
\label{sec:robustadvvsbenign}
\begin{figure}[t]
  \centering
  \includegraphics[width=0.4\textwidth]{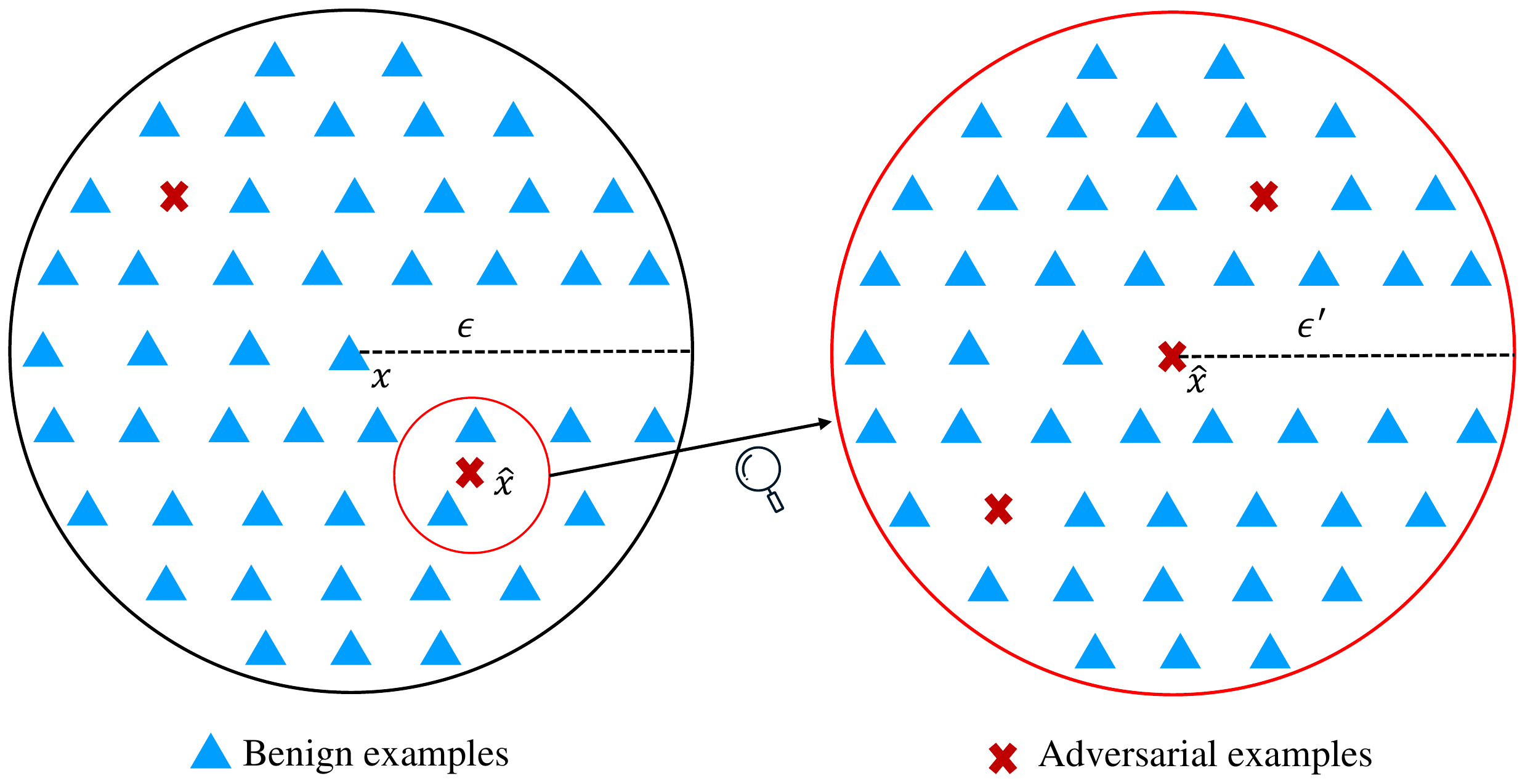}
  \vspace*{-3mm}
  \caption{An illustration of robustness of adversarial and benign examples. The left-part
  depicts $\epsilon$-neighbourhood of a benign example $x$ and the right-part
  depicts $\epsilon$-neighbourhood of the adversarial example $\hat{x}$, where
  each triangle denotes an example that can be correctly classified into the label $c_x$ and
  each cross denotes an example that cannot be correctly classified into the label $c_x$.}
  \label{fig:localrobustnessadvvsbenign}
\vspace*{-4mm}
\end{figure}

Our detection approach is based on the observation that
adversarial examples are much less robust than benign ones.
To understand the underlying reason, we briefly recap the processes of DL model training and adversarial example generation.
Training a DL model typically takes multiple epochs.
For each epoch, the training dataset is partitioned into multiple batches and each batch of input examples
is trained once. After each batch, the parameters are updated, e.g., via stochastic gradient descent.
Once all the epochs finish,
the DL model is ready for testing.
During training, each example in the dataset goes through
a number of epochs. Consequently,
it is often the case that the trained DL model achieves good generalization results.
Therefore, as illustrated in Figure~\ref{fig:localrobustnessadvvsbenign} (left-part),
under a reasonable distance threshold $\epsilon$,
most examples in the $\epsilon$-neighborhood of a benign example $x$ are also
benign while adversarial examples are relatively far away from the benign example $x$.

In the process of adversarial example generation,
the attacker crafts a perturbation $\Delta x$ for a benign example $x$
such that the resulting example $\Hat{x}=x+\Delta x$ is adversarial.
During the generation of adversarial examples, attackers often neglect robustness
due to the pursuit of other attributes,
such as minimal perturbation, invisibility, target label classification and query efficiency.
These attributes and robustness are often incompatible, and thus difficult to achieve simultaneously.
Consequently, adversarial examples with small distortion are very close to the decision boundary~\cite{wang2019adversarial} and
most examples in the $\epsilon'$-neighbourhood of the adversarial example $\hat{x}$ are
benign (with respect to the original example $x$).   
This is illustrated in Figure~\ref{fig:localrobustnessadvvsbenign} (right-part),
which is the zoom in of the $\epsilon'$-neighbourhood of the adversarial example $\hat{x}$ in Figure~\ref{fig:localrobustnessadvvsbenign} (left-part).

The above observation can be quantified using robustness.
We conduct a quantitative robustness comparison of benign and adversarial examples using the CLEVER score
on MNIST and CIFAR10 datasets.
For each dataset, we choose the first 100 images from the test dataset as subjects.
Adversarial examples are generated by applying
four representative attacks: FGSM, BIM, JSMA, and C\&W.
All the models and attack tools are taken from~\cite{feinman2017detecting,LID},
where the parameters are presented in
Table~\ref{table:env1-attack} in the supplementary material.
The CLEVER scores, in the form of a confidence interval of 90\% significance level~\cite{dekking2005modern}, are shown in Table~\ref{table:clever}.
Column \emph{Label for CLEVER} shows the label type for computing the CLEVER scores, where
Target-2/5 denotes the top-2/5 label of the example,
and LLC (least likely class) is the label with
the smallest probability.
Column \emph{Benign examples} shows the CLEVER scores of the benign examples.
Columns \emph{FGSM}, \emph{BIM}, \emph{JSMA} and \emph{C\&W} show the CLEVER scores of the adversarial examples
generated by  {FGSM},  {BIM},  {JSMA} and  {C\&W}, respectively.
Columns $\lambda$ show the ratio of the CLEVER scores of benign examples to
that of the adversarial ones for each attack.
The last column \emph{Avg. $\lambda$}  shows the ratio of
the CLEVER scores of benign examples to that of all the adversarial examples.

\begin{table*}[t]
\centering
\caption{CLEVER scores with confidence interval of 90\% significance level}
\vspace*{-3mm}
\label{table:clever}
\resizebox{1\textwidth}{!}{\begin{tabular}{c|c|c|cc| cc| cc|cc|c}
\hline
\multirow{2}{*}{Dataset} &
Label for &
\multirow{2}{*}{Benign examples} &
\multicolumn{8}{c|}{Adversarial examples} &
Avg. 
\\ \cline{4-11}
& CLEVER& & FGSM & $\lambda$ & BIM & $\lambda$ & JSMA & $\lambda$ & C\&W & $\lambda$  & $\lambda$\\ \hline 
\multirow{4}{*}{MNIST}   & Untarget & 3.5572 $\pm$ 0.3342 & 0.1093 $\pm$ 0.0506 & 32.55 & 0.0256 $\pm$ 0.0031 & 138.95 & 0.0550 $\pm$ 0.0060 & 64.68 & 0.0004 $\pm$ 0.0001 & 8893 & 74.77
\\ \cline{2-12}
& Target-2  & 3.6711 $\pm$ 0.3296   & 0.1148 $\pm$ 0.0427 & 31.98 & 0.0258 $\pm$ 0.0031 & 142.29 & 0.0558 $\pm$ 0.0063  & 65.79 & 0.0004 $\pm$ 0.0001 & 9178 & 74.62
\\ \cline{2-12}
& Target-5  & 3.8303 $\pm$ 0.3113 & 0.2047 $\pm$ 0.0431 & 18.71 & 0.1582 $\pm$ 0.0084 & 24.21 & 0.1898 $\pm$ 0.0096 & 20.18 & 0.1384 $\pm$ 0.0043 & 27.68 & 22.17
\\ \cline{2-12}
& LLC  & 3.8372 $\pm$ 0.3097 & 0.2390 $\pm$ 0.0421 & 16.06 & 0.1647 $\pm$ 0.0071 & 23.30 & 0.2120 $\pm$ 0.0076 & 18.10 & 0.1406 $\pm$ 0.0045 & 27.29 & 20.29
\\ \hline
\multirow{4}{*}{CIFAR10} & Untarget & 0.3851 $\pm$ 0.1850 & 0.2743 $\pm$ 0.1627 & 1.40 & 0.0329  $\pm$ 0.0033 & 11.71 & 0.0128 $\pm$ 0.0021 & 30.09 & 0.0005 $\pm$ 0.0002 & 770 & 4.81
\\ \cline{2-12}
& Target-2  & 0.4141 $\pm$ 0.1806 & 0.2971 $\pm$ 0.1675 & 1.39 & 0.0380 $\pm$ 0.0044 & 10.90 & 0.0129 $\pm$ 0.0021 & 32.10 & 0.0005 $\pm$ 0.0002 & 828 & 4.75
\\ \cline{2-12}
& Target-5  & 0.4657 $\pm$ 0.1913  & 0.3389 $\pm$ 0.1675  & 1.37 & 0.0971 $\pm$ 0.0117 & 4.80 & 0.0610 $\pm$ 0.0061 & 7.63 & 0.0925 $\pm$ 0.0168 & 5.03 & 3.16
\\ \cline{2-12}
& LLC  & 0.4829 $\pm$ 0.1913 & 0.3572 $\pm$ 0.1713 & 1.35 &  0.1091 $\pm$ 0.0132 & 4.43 & 0.0918 $\pm$ 0.0095 & 5.26 & 0.1035 $\pm$ 0.0180 & 4.67 & 2.92
\\ \hline
\end{tabular}}
\vspace*{-4mm}
\end{table*}


We can observe that the clever scores of benign examples are much larger than
that of adversarial ones for both MNIST and CIFAR10 datasets, though the difference varies from attacks and datasets.
This indicates that the difference of robustness between adversarial and benign examples is significant,
thus confirms our observation.
%
We also observe that the ratios $\lambda$ using untarget/target-2 label
for computing the CLEVER scores are larger than the ones using other labels.
This is because that the label of untarget or target-2 for each adversarial example is often the label of its benign counterpart,
which also confirms our observation that most examples in $\epsilon$-neighbourhood
of each adversarial example have the same label of the benign counterpart.

\subsection{Attack Cost: Adversarial vs. Benign}
\label{sec:attackcostadvvsbenign}

Based on the above observation,
one could design adversarial example detection approaches similar to other characterizations like label change rate~\cite{wang2019adversarial}. However,
existing techniques for robustness certification (with statistical guarantees) still have limited scalability, and hence are not able to handle large models efficiently.
For instance, on a single GTX 1080 GPU,
the cost of computing the CLEVER score is near:
  450 seconds for each MNIST example and 1150 seconds for each CIFAR10 example using untarget,
 50 seconds for each MNIST example and 128 seconds for each CIFAR10 example using target-2/5.

To effectively and efficiently detect adversarial examples,
we propose a novel detection approach, named \emph{attack as defense} (\tool for short),
which uses the cost of attacking an example to test its robustness.
The underlying assumption is that the more robust the example is, the more
difficult (a larger attack cost) it is to attack.
The implication is that
we can decide whether an input example is likely to be adversarial
by utilizing off-the-shelf attacks.


To leverage attack cost to detect adversarial examples,
the first problem needs to be tackled is
\emph{how to select attacks for defense}.
In general, the attack cost should be able to be quantified and reflect inputs' robustness.
As a result, FGSM is not suitable since it simply performs one-step  perturbation.
In contrast, iterative attacks (such as  BIM, JSMA, and C\&W) that iteratively search for adversarial examples with least distortion
could be leveraged, as the costs of such attacks can be quantified and are relevant to inputs' robustness.
%

\begin{figure}[t]
  \centering
  \subfigure[Score vs. time on MNIST]{\label{fig:scorevsattacktimeMNIST}
    \includegraphics[width=0.225\textwidth]{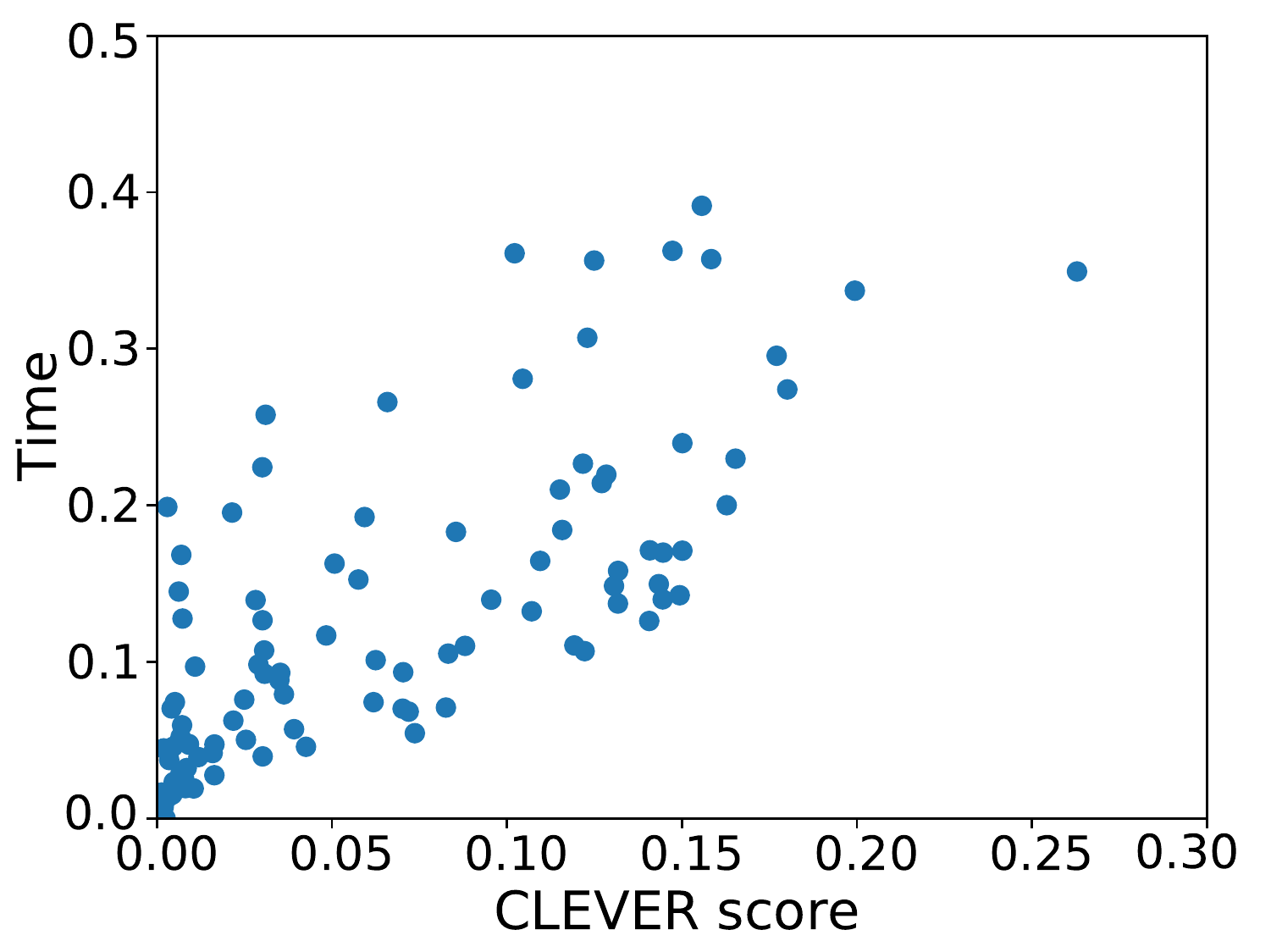}
    }
  \subfigure[Score vs. time on CIFAR10]{\label{fig:scorevsattacktimeCIFAR}
    \includegraphics[width=0.225\textwidth]{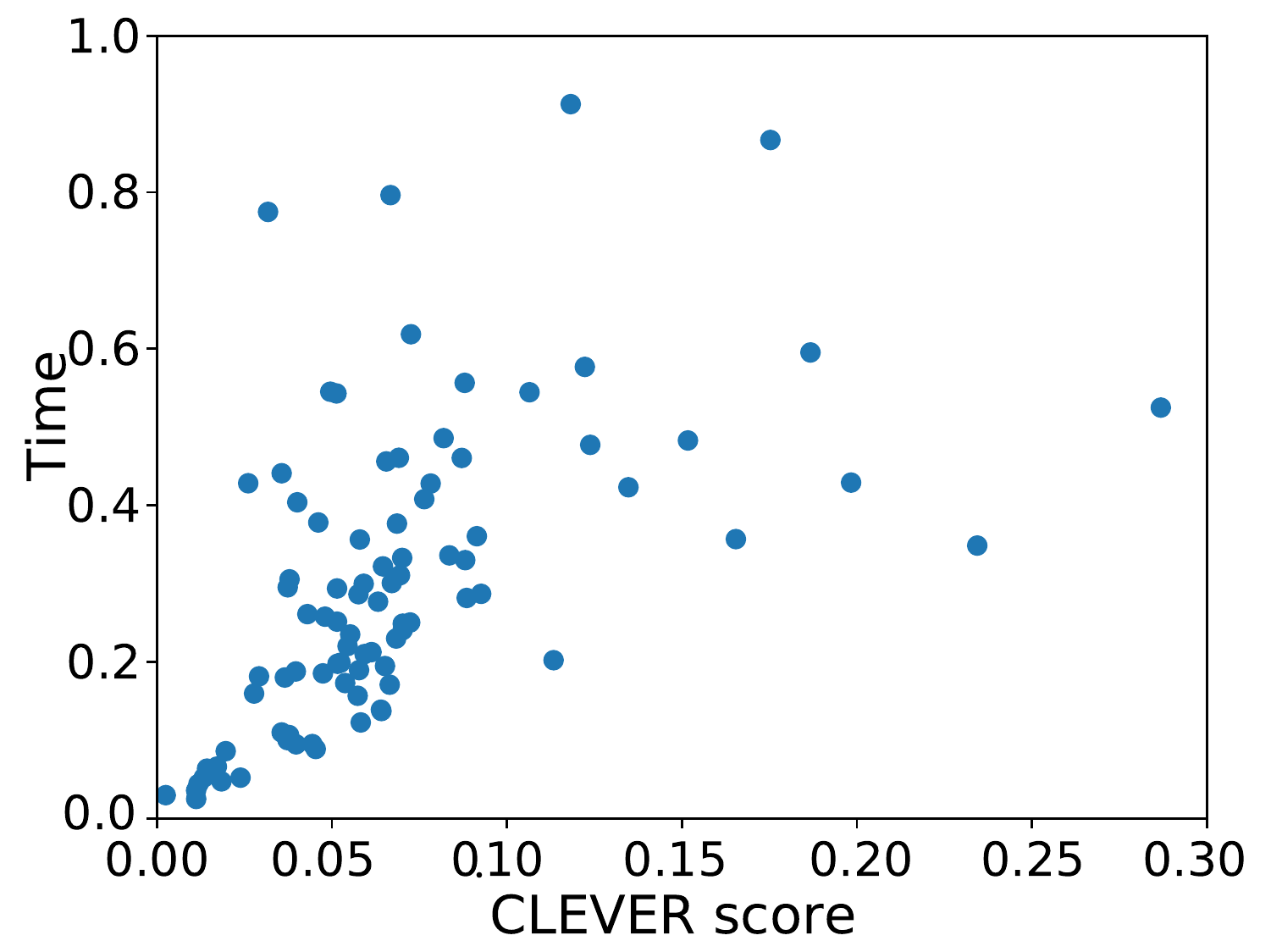}
    }
    \vspace*{-4mm}
  \caption{CLEVER score vs. attack time using JSMA}
  \label{fig:scorevsattacktime}
  \vspace*{-5mm}
\end{figure}

We illustrate this observation using JSMA. JSMA 
calculates a saliency map based on the Jacobian matrix to model the impact that each pixel imposes on the classification result.
During each iteration,
JSMA uses a greedy algorithm that modifies certain pixels to
increase the probability of the target label.
The process is repeated until finding an adversarial example
or reaching the termination criteria. The attack cost (time and iteration) of JSMA
depends on the robustness of each example. For an example $x$ that is less robust than another one $x'$,
an adversarial example of $x$ can be quickly constructed using less time/iteration than the one of $x'$.
To further test this observation,
we compare the attack time of JSMA on 100 MNIST and 100 CIFAR10 examples whose CLEVER scores range from $0$ to $0.3$.
The results are reported as scatter plots in
Figure~\ref{fig:scorevsattacktimeMNIST} and Figure~\ref{fig:scorevsattacktimeCIFAR}, which confirm
our observation.



\begin{figure}[t]
\centering
\includegraphics[width=0.45\textwidth]{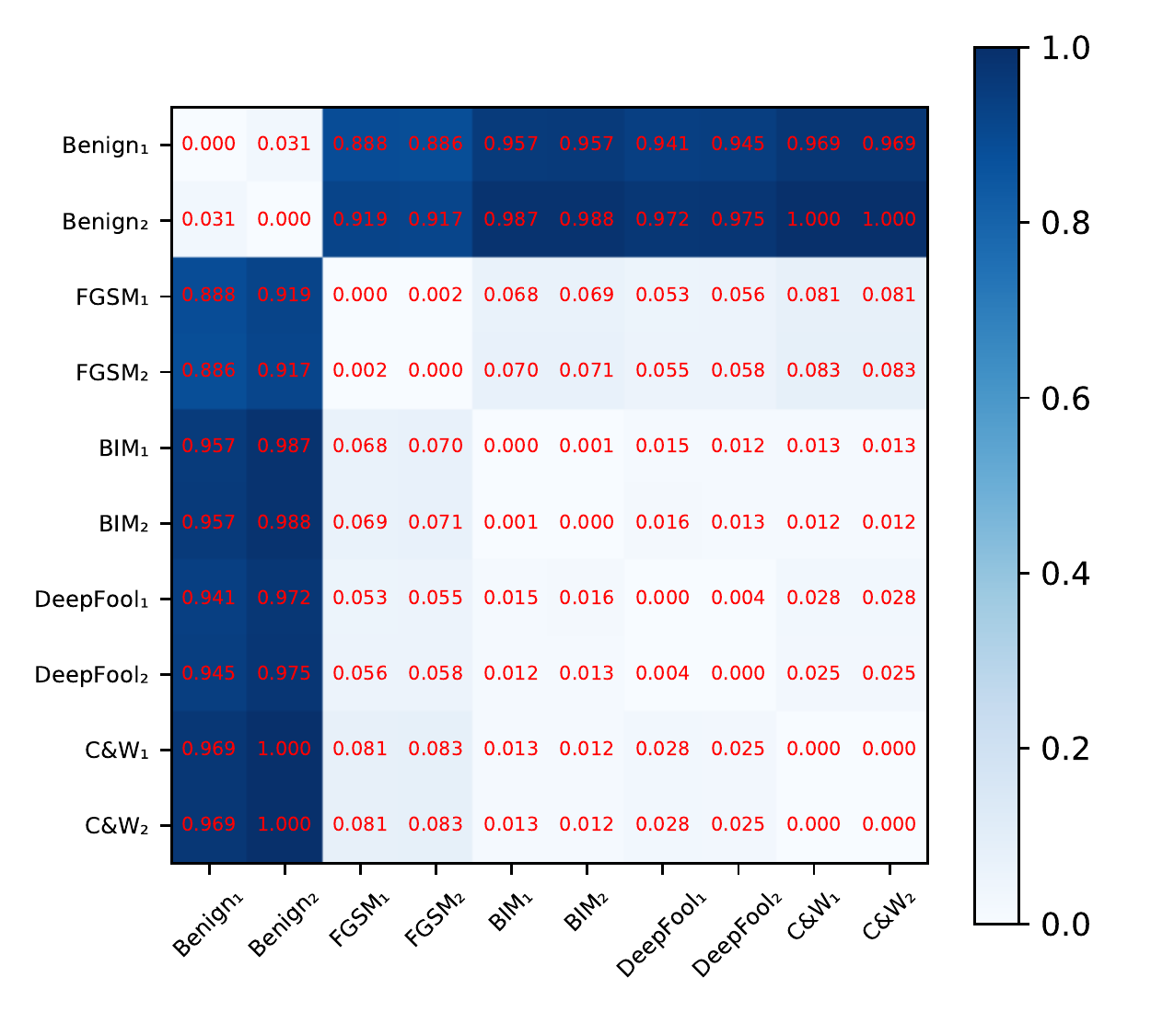}
\vspace*{-6mm}
\caption{Euclidean distances of the average number of iterations between each pair of sets of examples}
\label{fig:distance}
\vspace*{-3mm}
\end{figure}

The next problem is then \emph{how to characterize attack costs for different kinds of attacks}.
The most direct indicator of attack costs is the attack time as demonstrated in Figure~\ref{fig:scorevsattacktime}.
The attack time of different examples can reflect their robustness.
However, the attack time is easily affected by the real-time performance of computing devices in a physical environment,
which makes the variance of attack time intolerable.
Therefore, in this work, we propose to use the number of iterations of the attacks as the indicator of the attack costs.
For an iterative attack, the number of iterations is positively correlated with attack time (cf. the supplementary material). 

To demonstrate the effectiveness of the attack costs for
characterizing adversarial examples,
we choose 5 types of images (Benign, FGSM, BIM, DeepFool and C\&W),
each of which randomly select 1,000 samples,
and divided each type of images into two independent sets.
Then we use JSMA to attack these images and record the number of iterations required.
To show the difference in attack costs,
we calculate the average Euclidean distance of the number of iterations
between each pair of sets of examples,
the results are presented in Figure~\ref{fig:distance}.
We can see that for different types of examples (adversarial vs. benign),
the distance is enormous.
While for the same types of examples (adversarial vs. adversarial or benign vs. benign),
the distance is close to zero.
It is worth mentioning that for the examples generated by different attacks,
the distance is also very similar,
meaning that even if the adversarial examples are generated by different attacks,
they are also ``cognate" examples and have similar attack costs.


Utilizing the diversity of attack methods,
an ensemble detection method can be constructed jointly.
As we mentioned before,
iterative attacks have the potential to be used as defense,
so we can integrate multiple iterative attacks to
derive a more robust defense.
Different attacks have the ability to capture different characterizations.
For example, JSMA crafts adversarial examples based on the $L_0$ norm,
while BIM is based on the $L_\infty$ norm,
thus they measure the robustness of inputs
under different distance metrics.
Thanks to various types of attacks,
ensemble multiple attacks will make the defense more reliable
and difficult to bypass.



\section{Detection Approach}
In this section, we consider how to detect adversarial examples by leveraging attack costs.
In this work, we propose two effective detection approaches that are based on
$k$-nearest neighbors (K-NN) and standard score (Z-score), respectively.
The former requires both benign and adversarial examples, while the latter
requires only benign examples.

Hereafter, we sometimes denote by attack$_d$ the attack that is used as defense, i.e., to generate attack costs.

\subsection{K-NN based Detection Approach}
Assume that we have two disjoint sets:
$B$ the set of benign examples
and $A$ the set of adversarial examples.

\smallskip
\noindent \textbf{Single detector.} Let us consider the attack$_d$ $o$.
The attack cost $\alpha_y$ of attacking an example $y$ using the attack$_d$ $o$ is regarded as the {fingerprint} of $y$.
We can generate
a set of fingerprints $\{\alpha_y\mid y\in A\cup B\}$ from the examples $y\in A\cup B$ by utilizing the attack$_d$ $o$.
For each unknown input $x$ and parameter $K$,
we first compute the attack cost $\alpha_x$ of the input $x$ using the attack$_d$ $o$
and then identify the $K$-nearest neighbors $N_K=\{\alpha_{y_i}\mid 1\leq i \leq K\}$ of $\alpha_x$ from the set $\{\alpha_y\mid y\in A\cup B\}$.
The set $N_k$ is partitioned into two subsets:
$A_x=\{y\in A\mid \alpha_y\in N_K\}$ and $B_x=\{y\in B\mid \alpha_y\in N_K\}$.
The input $x$ is classified as adversarial if $|A_x|>|B_x|$,
namely, the number of adversarial examples is larger than that of benign ones
in K-neighbourhood of the input $x$.

\smallskip
\noindent \textbf{Ensemble detector.} The K-NN based detection approach can be easily generalized from one attack$_d$ to multiply attacks$_d$
$o_1,\cdots,o_n$, leading to a more robust detector.
Under this setting, the fingerprint of an example $y$
is a vector of attack costs, $\vec{\alpha}_y=(\alpha_y^1,\cdots, \alpha_y^n)$,
where for every $1\leq j\leq n$, $\alpha_y^j$
is the attack cost of the example $y$ by utilizing the attack$_d$ $o_j$.
Consequently, we can generate
a set of fingerprints $\{\vec{\alpha}_y\mid y\in A\cup B\}$ from the examples $y\in A\cup B$ by utilizing the attacks$_d$ $o_1,\cdots,o_n$.
Similar to the single attack setting,
for each unknown input $x$ and parameter $K$, we identify the
$K$-nearest neighbors $N_K=\{\vec{\alpha}_{y_i}\mid 1\leq i \leq K\}$ of
the fingerprint $\vec{\alpha}_x$ of the input $x$
and partition $N_k$ into two subsets:
$A_x=\{y\in A\mid \vec{\alpha}_y\in N_K\}$ and $B_x=\{y\in B\mid \vec{\alpha}_y\in N_K\}$.
The input $x$ is classified as adversarial if $|A_x|>|B_x|$.

\begin{figure*}
\centering
\subfigure[FGSM$_d$]{
\includegraphics[width=0.12\textwidth, height=3cm]{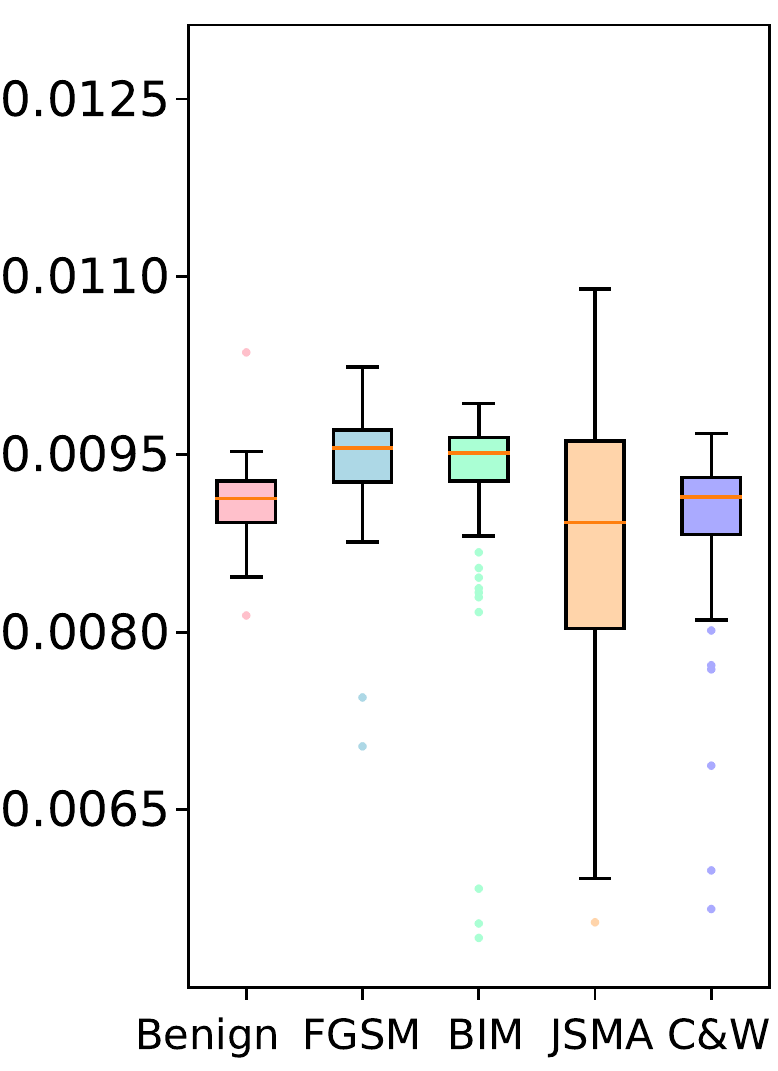}
}\hspace{-2mm}
\subfigure[BIM$_d$]{
\includegraphics[width=0.12\textwidth, height=3cm]{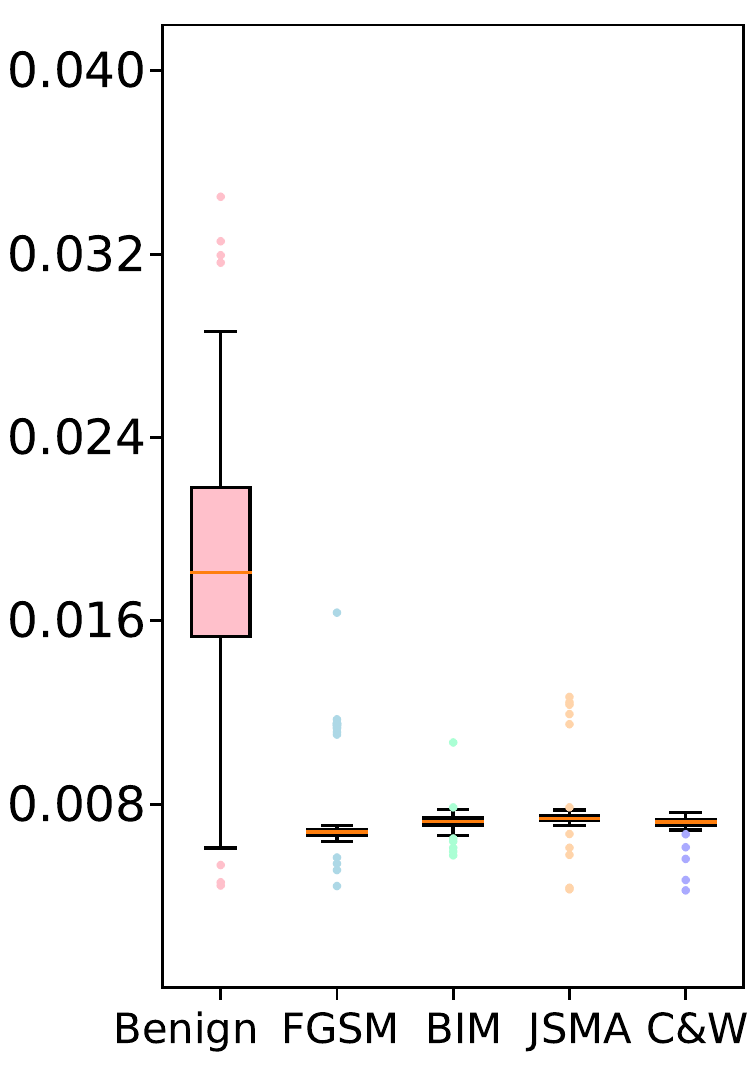}
}\hspace{-2mm}
\subfigure[BIM2$_d$]{
\includegraphics[width=0.12\textwidth, height=3cm]{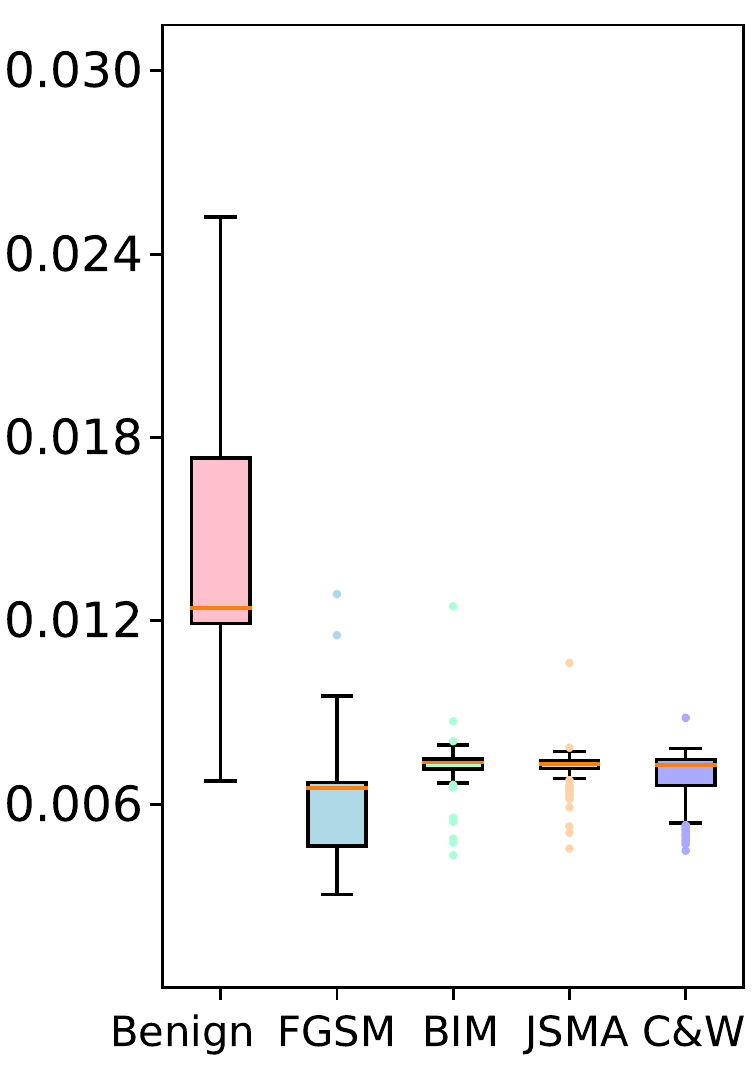}
}\hspace{-2mm}
\subfigure[JSMA$_d$]{
\includegraphics[width=0.12\textwidth, height=3cm]{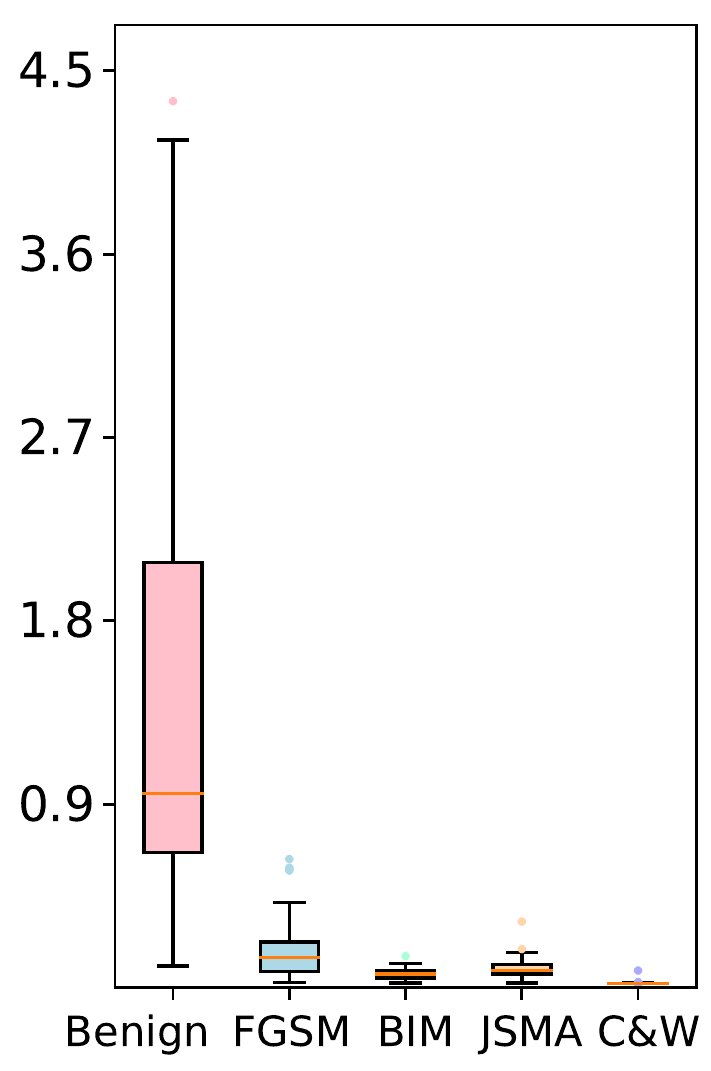}
}\hspace{-2mm}
\subfigure[C\&W$_d$]{
\includegraphics[width=0.12\textwidth, height=3cm]{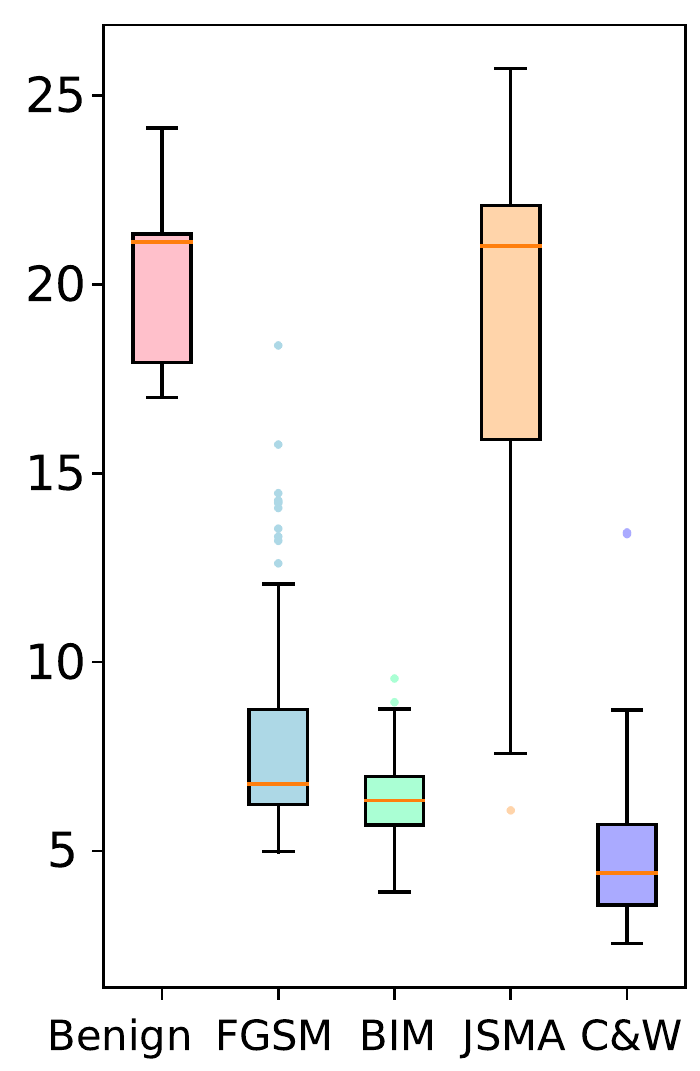}
}\hspace{-2mm}
\subfigure[L-BFGS$_d$]{
\includegraphics[width=0.12\textwidth, height=3cm]{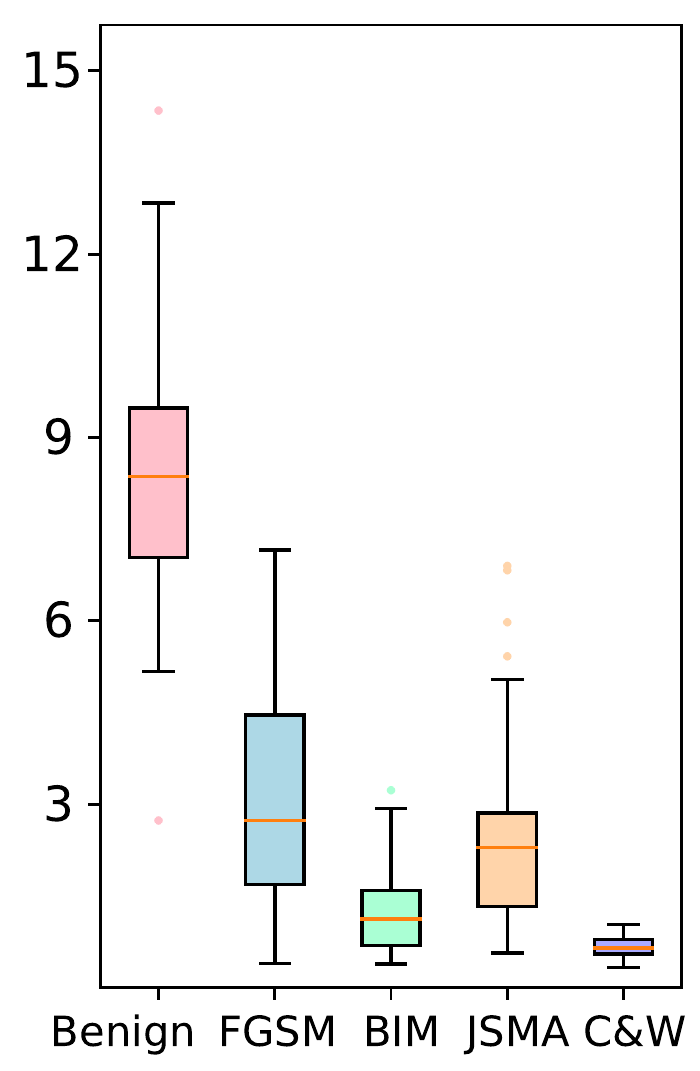}
}\hspace{-2mm}
\subfigure[LSA$_d$]{
\includegraphics[width=0.12\textwidth, height=3cm]{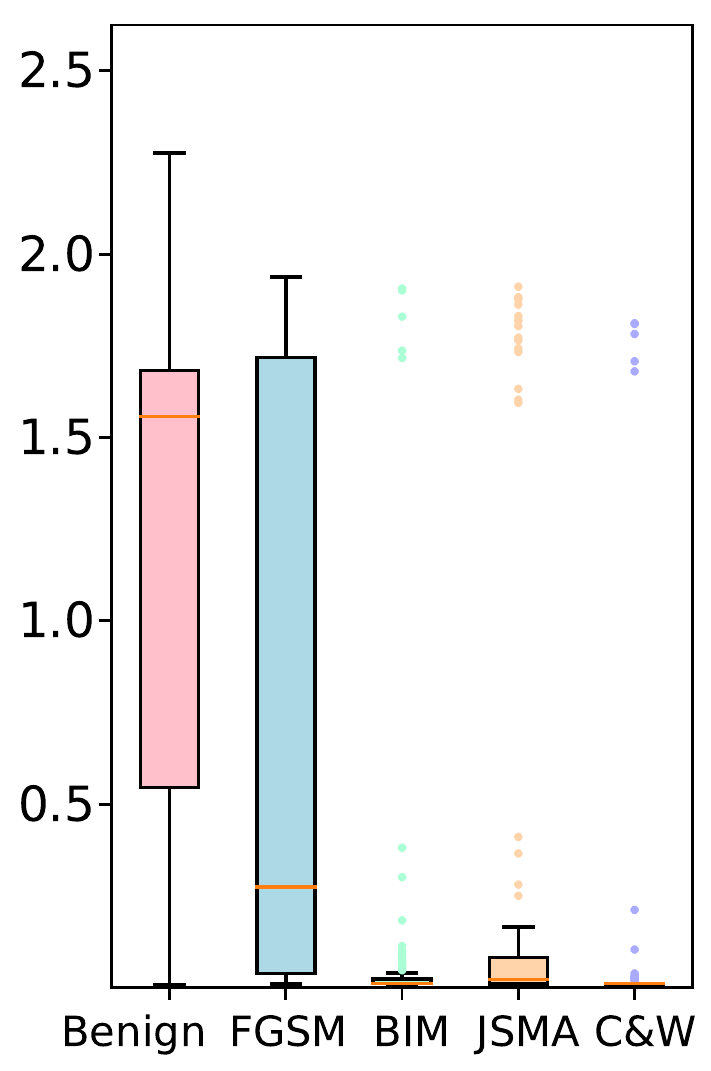}
}\hspace{-2mm}
\subfigure[DBA$_d$]{
\includegraphics[width=0.12\textwidth, height=3cm]{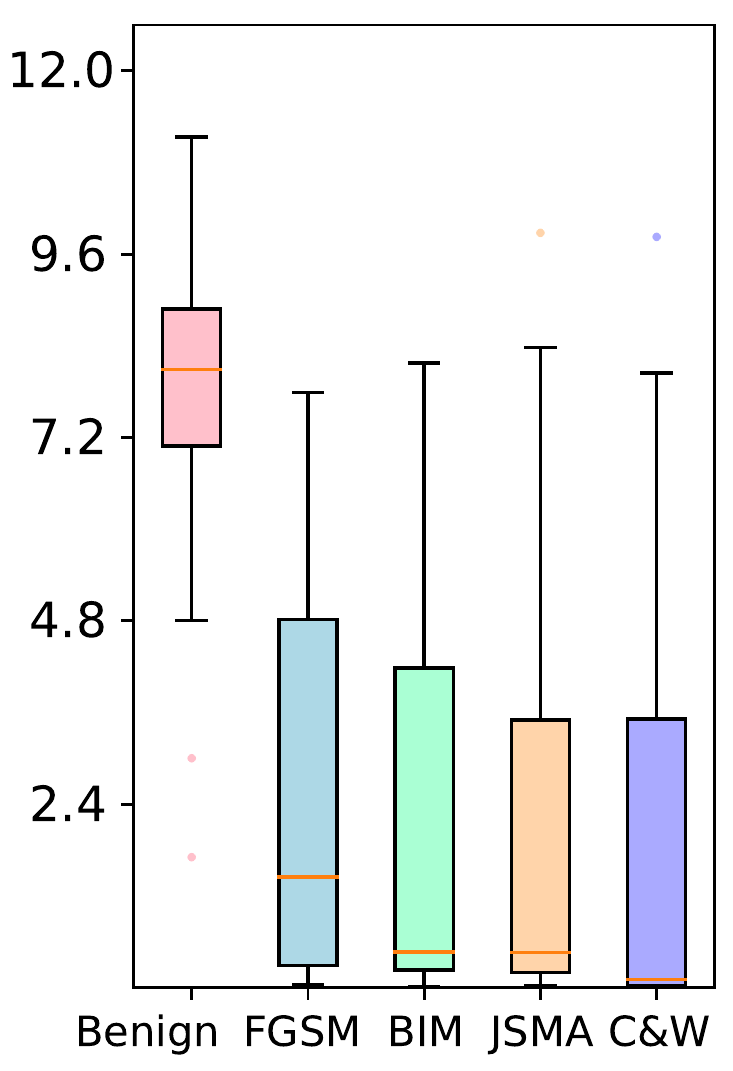}
}
\vspace*{-4mm}
\caption{Attack time of benign and adversarial examples, where $y$-axis means seconds} 
\label{fig:time-mnist}
\vspace*{-3mm}
\end{figure*}

\subsection{Z-Score based Detection Approach}
Z-score is a well-known concept in statistics
for measuring a sample in terms of its relationship to the mean and standard deviation of a dataset~\cite{LM11}.
The Z-score of a sample $i$ is defined by:
$z=\frac{i-\mu}{\sigma}$,
where $\mu$ is the sample mean and $\sigma$ is the sample standard
deviation.
Intuitively, the score $z$ indicates how many standard
deviations that the sample $i$ is far away from the sample mean.
Our Z-Score based detection approach leverages the distribution of attack costs of benign examples
to check whether an example $y$ is adversarial or not.
Thus, it is likely more robust with respect to unseen attacks.

\smallskip
\noindent \textbf{Single detector.} Let us consider a set $B$ of benign examples and an attack$_d$ $o$ as defense.
We can compute the distribution of attack costs of the examples in $B$.
Assume that the distribution is an approximately normal distribution $N(\mu,\sigma^2)$. 
Otherwise, we can transform it by applying the Box-Cox power transformation~\cite{boxcox}.
Thus, the Z-score $z_y$ of an example $y$ is defined as:
$z_y=\frac{\alpha_y-\mu}{\sigma}$, where $\alpha_y$ denotes the  cost of attacking
$y$ using the attack$_d$ $o$.
For a given ratio $h$ of the sample standard
deviation as the threshold,
based on our observation that adversarial examples are less robust than benign ones,
an input $x$ is classified to adversarial
if $z_x<h$,
i.e., $x$ is $h$ standard deviations away from the sample mean.

\smallskip
\noindent \textbf{Ensemble detector.}
We can also generalize this approach from one attack$_d$ to multiply attacks$_d$
$o_1,\cdots,o_n$.
For each attack$_d$ $o_j$, we can construct a Z-Score based detector $d_j$,
resulting in detectors $d_1,\cdots,d_n$.
The ensemble detector determines whether an input is adversarial or not
by taking into account the results of all the detectors $d_1,\cdots,d_n$.
Consider $k\leq n$,
the ensemble detector classifies an input to benign if $k$ detectors classify the input to benign, otherwise adversarial.
The ensemble detector would have high true positive rates when $k=n$,
high true negative rates when $k=1$.

\begin{figure}[t]
\centering
\subfigure[BIM$_d$]{
\includegraphics[width=0.115\textwidth, height=3cm]{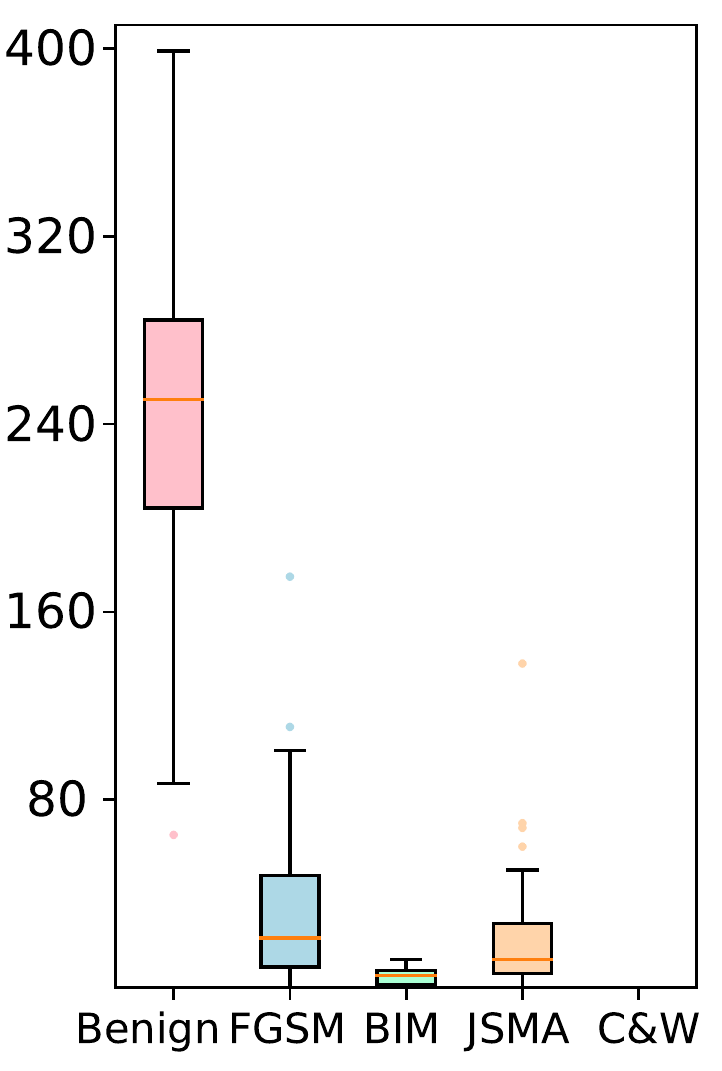}
}\hspace{-2mm}
\subfigure[BIM2$_d$]{
\includegraphics[width=0.115\textwidth, height=3cm]{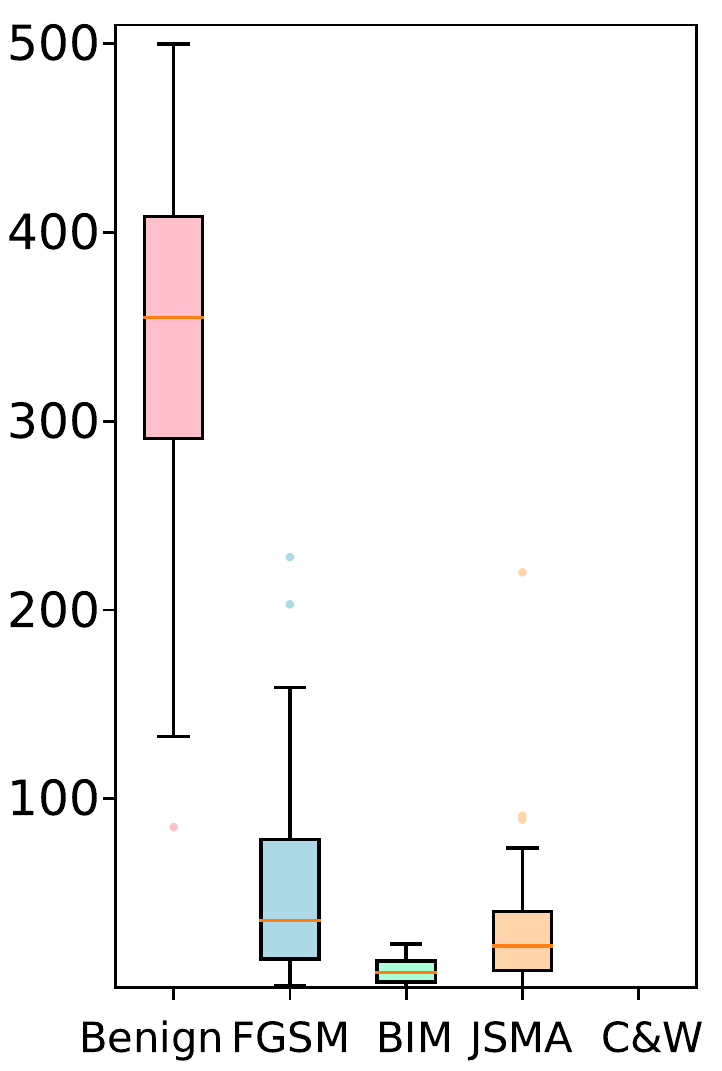}
}\hspace{-2mm}
\subfigure[JSMA$_d$]{
\includegraphics[width=0.115\textwidth, height=3cm]{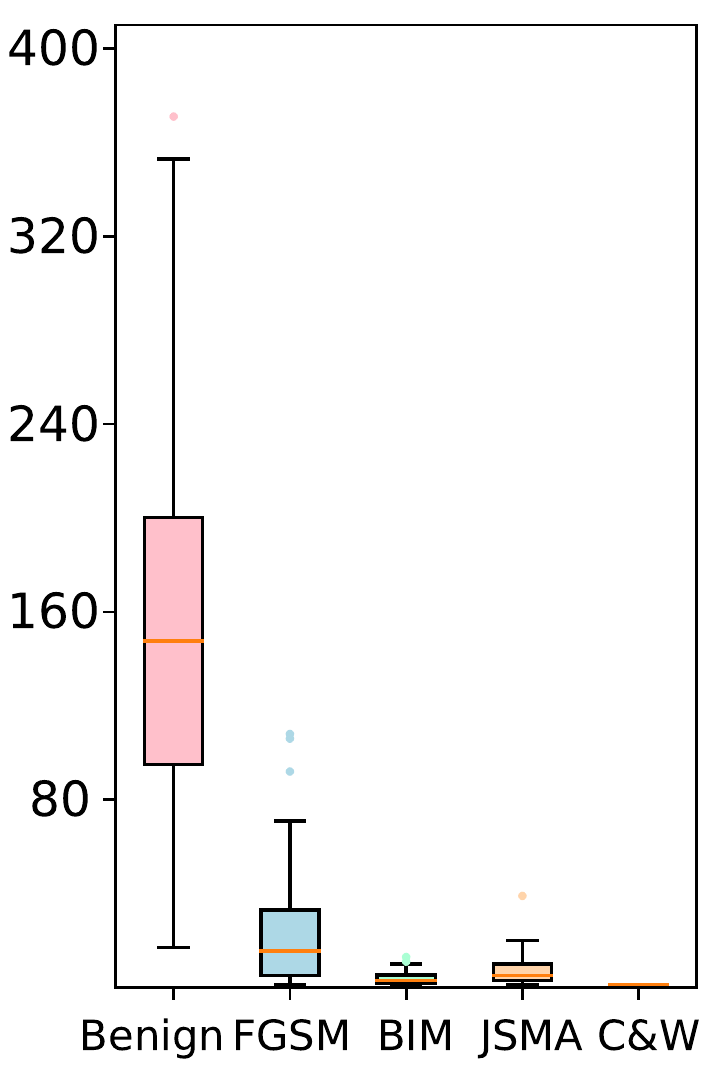}
}\hspace{-2mm}
\subfigure[DBA$_d$]{
\includegraphics[width=0.115\textwidth, height=3cm]{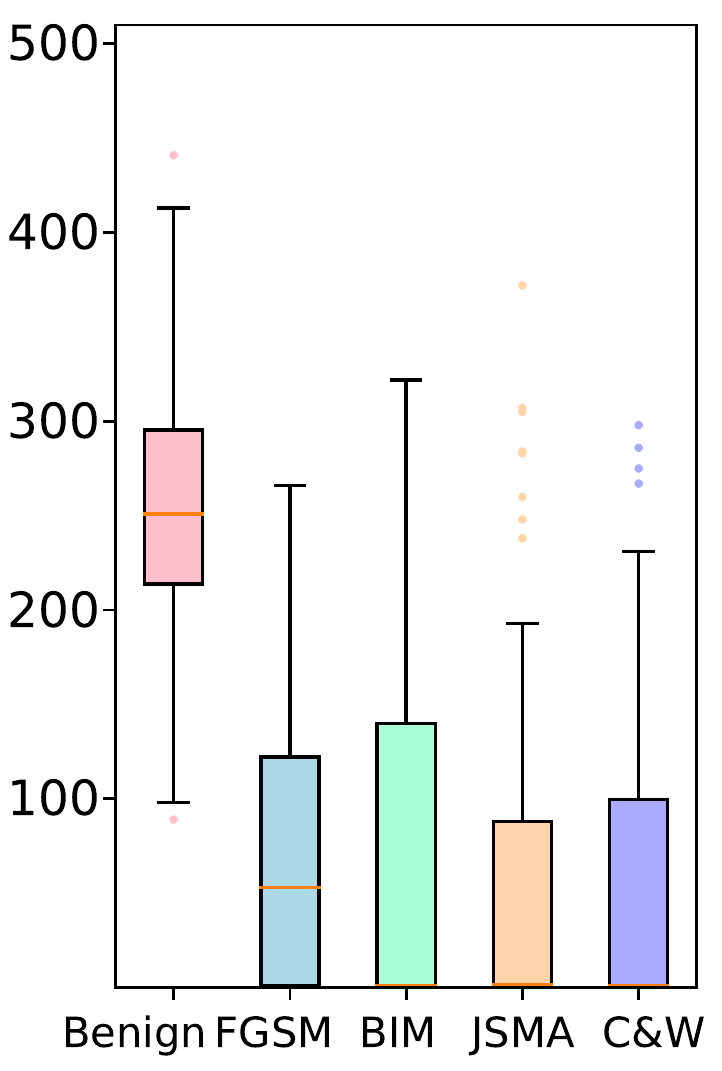}
}
\vspace*{-5mm}
\caption{Attack iterations comparison} 
\label{fig:iter-mnist}
\vspace*{-4mm}
\end{figure}

\section{Evaluation}
\label{sec:imple}
In this section, we evaluate our approach on three widely used
datasets MNIST, CIFAR10 and ImageNet.
The experiments are designed to answer the following
research questions:
%
\begin{enumerate}[label=\textbf{RQ\arabic*. },itemindent=*,leftmargin=*,itemsep=0pt]
    \item How to select effective attacks for defense?
    \item How effective are the selected attacks for defense?
    \item How effective and efficient is \tool (i.e., detection)?
\end{enumerate}

\subsection{{\bf RQ1:} Attack Selection}
We answer {\bf RQ1} by comparing attack costs of adversarial and benign examples.
Adversarial examples are crafted in the same setting as in Section~\ref{sec:robustadvvsbenign}, using the first 1,000 images from MNIST. 

For defense, we choose eight attacks:
FGSM, BIM, BIM2 (BIM under $L_2$ norm), JSMA, C\&W, L-BFGS, LSA, DBA
from Foolbox~\cite{RBB17}, recommended by~\cite{TCBM20}.
For ease of reading,
an attack used as a defense is marked by a subscript $d$,
e.g., FGSM$_d$.
According to the results of Table~\ref{table:clever},
we use untarget attack as defense,
unless the attack only supports target attack
for which we use target-2 attack.
We use the default parameters of Foolbox
except that BIM$_d$ and BIM2$_d$ immediately terminate
when an adversarial example is found,
and DBA$_d$ terminates when the MSE between the adversarial example and its original version is less than 0.02, 
otherwise the number of iterations is fixed, namely, the attack costs
of all the examples will be similar.

The results in terms of attack time
are reported as box plots in Figures~\ref{fig:time-mnist}.
It is not surprising that the attack time of adversarial and benign examples using FGMS$_d$ are similar,
as it is a one-step attack.
Though the results show the variation
between different attacks and defenses,
the differences are often significant
when an iterative attack is
used as defense, e.g., BIM$_d$, BIM2$_d$, JSMA$_d$, L-BFGS$_d$ and DBA$_d$. We find that the differences are not stable when C\&W$_d$ and LSA$_d$ are used.
This is because that C\&W$_d$ implements a binary search to minimize distortion
and may stop searching when there is a bottleneck,
while LSA$_d$ suffers from the lower attack success rate
which causes it to have many outliers.

Based on the above results, considering the efficiency and discrepancy in attack time
between benign and adversarial examples,
BIM$_d$, BIM2$_d$, JSMA$_d$, and DBA$_d$
will be used as defense in the follow-up experiments.
These four methods cover both white-box and black-box attacks,
as well as different distance metrics $L_0$, $L_2$ and $L_\infty$.
It should be noted that with an amount of adversarial attacks being proposed
(more than 2,000 papers in 2020),
it is impossible to evaluate all of them.
In this work, we only analyze the pros and cons of the above methods
and choose suitable attacks. 

The above experiments demonstrate how to quickly select a suitable attack as defense.
In order to ensure the reliability of BIM$_d$, BIM2$_d$, JSMA$_d$, and DBA$_d$,
we also analyze the numbers of iterations. The results are reported as box plots in Figure~\ref{fig:iter-mnist},
where the maximal number of iterations of BIM$_d$ and BIM2$_d$ is increased from
the default value 10 to 500 in order to obtain a more significant difference.
Remark that we did not tune parameters,
these widely used parameters are sufficient to achieve expected results.
Fine-tuning parameters may yield better results.
We can observe that the differences in the numbers of iterations are consistent with that of
attack time, therefore. Since the number of iterations does not depend on computing devices, thus, we will use the number of iterations
as the indicator of attack costs in the follow-up experiments.

\begin{tcolorbox}[left=0mm,right=0mm,top=0mm,bottom=0mm]
\textbf{Answer to RQ1:}
Both attack time and the number of iterations can be used to select effective attacks for defense,
while non-iterative attacks are not effective.
\end{tcolorbox}

\subsection{{\bf RQ2:} Effectiveness of Attacks as Defense}
We answer {\bf RQ2} by comparing our approach with four promising approaches as baselines.
The evaluation metric used here is AUROC
(area under the receiver operating characteristic curve),
which is one of the most important evaluation metrics for measuring the performance of classification indicators.
The large the AUROC, the better the approach for detecting adversarial examples.

The first one~\cite{feinman2017detecting}, denoted by BL$_1$,
uses a Gaussian Mixture Model to model outputs from the final hidden layer of a CNN,
which was considered to be \emph{the most effective defense on MNIST} among ten detections in~\cite{CW17a}.
The second one~\cite{LID}, denoted by BL$_2$,
uses local intrinsic dimension to represent and distinguish adversarial subspace and claims to be better than BL$_1$.
BL$_3$~\cite{wang2019adversarial}  uses label change rate through model mutation testing to distinguish adversarial examples.
BL$_4$~\cite{Wang2019Dissector} dissects the middle layer outputs to construct a fault tolerance approach.
We emphasize that BL$_3$ and BL$_4$~\cite{Wang2019Dissector} were published in ICSE'19 and ICSE'20, respectively.
BL$_2$ is implemented on BL$_1$ which
is implemented on Keras (called Env$_1$),
while BL$_3$ and BL$_4$ is implemented on Pytorch (called Env$_2$).
Since the performance of these four defenses may vary due to platforms, CNN models and attack settings,
for a fair comparison,
we implement our approach in both Env$_1$ and Env$_2$ and
conduct comparison directly using the same target models and attacks provided by each of them.
(Parameters are given in Tables~\ref{table:env1-attack} and~\ref{table:env2-attack} in the supplementary material.)


\begin{table}[t]
\centering
\caption{AUROC comparison of our approach over baselines}
\vspace*{-3mm}
\label{table:auroc-17}
\resizebox{0.48\textwidth}{!}{
\begin{tabular}{c|c|c|c|c|c||c|c}
\hline
Env$_1$    & Attack    & JSMA$_d$   & BIM$_d$
& BIM2$_d$ & DBA$_d$  & BL$_1$ & BL$_2$\\ \hline
\multirow{4}{*}{MNIST}
& FGSM     & 0.9653  & \textbf{0.9922} & 0.9883 & 0.9504 & 0.8267 & 0.9161
\\ \cline{2-8}
& BIM    & 0.9986  & \textbf{0.9996} & 0.9995 & 0.9625 & 0.9786 & 0.9695  \\
\cline{2-8}
& JSMA & \textbf{0.9923}  & 0.9922 & 0.9914 & 0.9497 & 0.9855 & 0.9656 \\
\cline{2-8}
& C\&W    & \textbf{1.0}  & \textbf{1.0} & \textbf{1.0} & 0.9672 & 0.9794 & 0.9502 \\
\hline
\multirow{4}{*}{CIFAR10}
& FGSM    & 0.6537 & 0.712 & 0.6474 & 0.6977 & 0.7015 & \textbf{0.7891}
\\ \cline{2-8}
& BIM    & 0.8558   & \textbf{0.8636} & 0.861 & 0.8276 & 0.8255 & 0.8496
\\ \cline{2-8}
& JSMA    & 0.9459 & \textbf{0.955}  & 0.9526 & 0.9452 & 0.8421 & 0.9475
\\ \cline{2-8}
& C\&W      & 0.9905 & 0.9984 & \textbf{0.9988} & 0.9833 & 0.9217 & 0.9799
\\ \hline \hline




Env$_2$    & Attack    & JSMA$_d$   & BIM$_d$
& BIM2$_d$  & DBA$_d$  & BL$_3$ & BL$_4$ \\ \hline
\multirow{5}{*}{MNIST}
& FGSM     & 0.9665  & \textbf{0.9883} & 0.9846 & 0.9595 & 0.9617 & -
\\ \cline{2-8}
& JSMA    & 0.9971  & \textbf{0.9984} & 0.9974 & 0.984 & 0.9941 & - \\
\cline{2-8}
& DeepFool & 0.9918  & \textbf{0.9971} & 0.9951 & 0.9587 & 0.9817 & - \\
\cline{2-8}
& C\&W & 0.9456  & \textbf{0.9870} & 0.9769 & 0.8672 & 0.9576 & -  \\
\cline{2-8}
& BB & 0.9746  & \textbf{0.9895} & 0.9852 & 0.9535 & 0.9677 & - \\
\hline
\multirow{4}{*}{CIFAR10}
& FGSM     & 0.8808 & 0.8994 & \textbf{0.8998} & 0.8746 & 0.8617 & -
\\ \cline{2-8}
& JSMA    & 0.9774  & \textbf{0.9890} & 0.9873 & 0.9566 & 0.9682 & -
\\ \cline{2-8}
& DeepFool & 0.9832 & 0.9898  & \textbf{0.9902} & 0.9769 & 0.9614
\\ \cline{2-8}
& C\&W  & 0.8842   & \textbf{0.9176}  & 0.9175 & 0.9004 & 0.9063 & -
\\
\hline

\multirow{4}{*}{ImageNet}
& FGSM     & 0.973  & 0.9763 & \textbf{0.9782} & 0.9625 & - & 0.9617
\\ \cline{2-8}
& JSMA    & \textbf{0.9962}  & 0.9805 & 0.99 & 0.9937 & - & 0.9695
\\ \cline{2-8}
& DeepFool & \textbf{0.9958}  & 0.9793 & 0.9892 & 0.9891 & - & 0.9924
\\ \cline{2-8}
& C\&W      & 0.9873  & 0.9731 & 0.9801 & \textbf{0.9924} & - & 0.9636
\\ \hline
\end{tabular}
}\vspace*{-3mm}
\end{table}

Table~\ref{table:auroc-17} shows the results in AUROC,
where the best one is highlighted in \textbf{bold} font.
Note that BL$_1$, BL$_2$ and BL$_3$ only support the
MNIST and CIFAR10 datasets, thus there are no results
on the ImageNet dataset. 
Though BL$_4$ considered all these datasets,
its open source tool only supports the ImageNet dataset when we conduct experiments.
Overall, we can observe that our approach outperforms the baselines in most cases.
It is worth mentioning that our defense parameters
are the same in both environments,
which shows its universality,
namely, users do not need to adjust parameters for a specific DL model or platform.

Among the four defenses JSMA$_d$, BIM$_d$, BIM2$_d$ and DBA$_d$ on MNIST and CIFAR10,
BIM$_d$ performs better than the others in almost all the cases,
while DBA$_d$ performs worse than the others in most cases.
This is due to that DBA$_d$ is a black-box attack which is less powerful
than the other white-box attacks. 
An interesting phenomenon is that
the AUROC of JSMA$_d$ and DBA$_d$ on ImageNet is better than on MNIST and CIFAR10.
This is because that for images with large dimension,
each perturbation generated by JSMA and DBA is smaller than that of BIM,
resulting in a fine-grained attack as well as a fine-grained indicator of examples' robustness.
One may find that BL$_2$ performs better than the others
on CIFAR10 adversarial examples crafted by FGSM.
This may be because that the performance of the model is
too poor as its accuracy is only 80.3\% on the testing dataset
(cf. Table~\ref{table:env1-attack} in the supplementary material).
Due to the poor performance of the CIFAR10 model,
most attacks of benign examples can be achieved easily,
hence the attack costs of adversarial examples generated by FGSM
are close to benign examples.
This problem could be alleviated by using state-of-the-art models (such as the model in Env$_2$)
or improving the robustness of the model
(such as adversarial training, cf. Section~\ref{sec:a2d+at}).



\begin{tcolorbox}[left=0mm,right=0mm,top=0mm,bottom=0mm]
\textbf{Answer to RQ2:}
Against most attacks on 2 popular platforms and 3 widely-used datasets, the selected white-box attacks JSMA$_d$, BIM$_d$ and BIM2$_d$ are more effective than
the baselines.
\end{tcolorbox}


\subsection{{\bf RQ3:} Effectiveness and Efficiency of \tool}
We answer {\bf RQ3} by applying our K-NN and Z-score based detectors
to check benign examples and adversarial examples generated by the attacks from BL$_3$ in Env$_2$. 
We do not consider other baselines, which only considered the results of AUROC
or do not provide cost analysis.
In order to avoid overfitting,
we selected different data as the training set and the test set.


\subsubsection{Effectiveness}
\noindent
{\bf K-NN based detectors.} For each dataset, each attack$_d$ of BIM$_d$, BIM2$_d$, JSMA$_d$ and DBA$_d$ as defense
and each attack $a$ in Env$_2$,
we construct a K-NN based detector through the attack costs of 1,000 benign examples and
1,000 attack $a$ crafted adversarial examples using the defense attack$_d$.
We also construct a K-NN based ensemble detector END, which consists of 1,000 benign examples and
1,000 adversarial examples, where each attack of Env$_2$ contributes 1000 / N adversarial examples (N is the number of attacks).
We set $K=100$. Results on tuning
$K$ and ratio between benign, adversarial examples and classification algorithms are given in Section~\ref{sec:parametertuning}
in the supplementary material.

%

%

\begin{figure}[t]
\centering
\subfigure[K-NN, MNIST]{\label{fig:detectorknnmnist}
\includegraphics[width=0.225\textwidth]{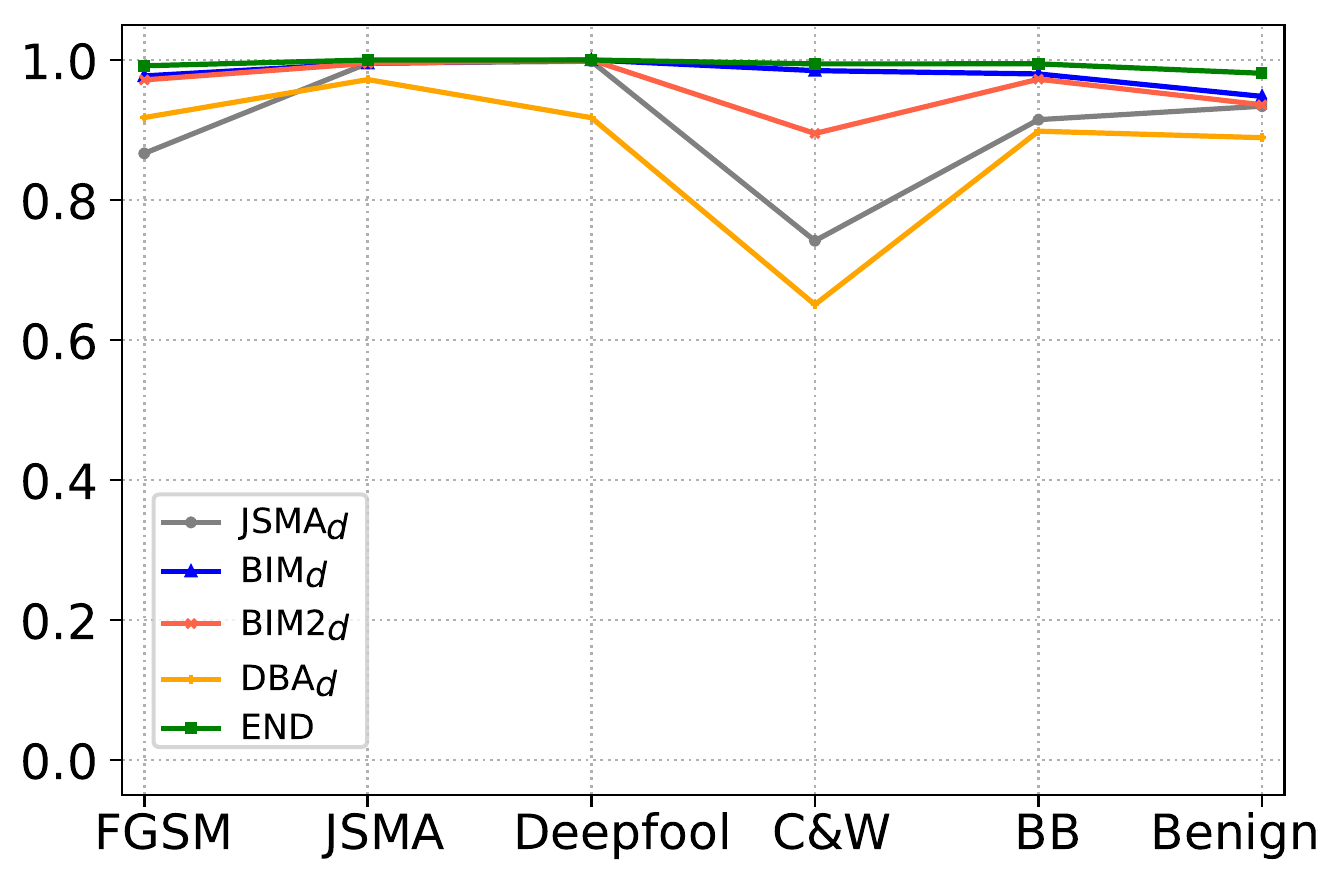}
}
\hspace*{-2mm}\vspace*{-2mm}
\subfigure[Z-score, MNIST]{\label{fig:detectorzscoremnist}
\includegraphics[width=0.225\textwidth]{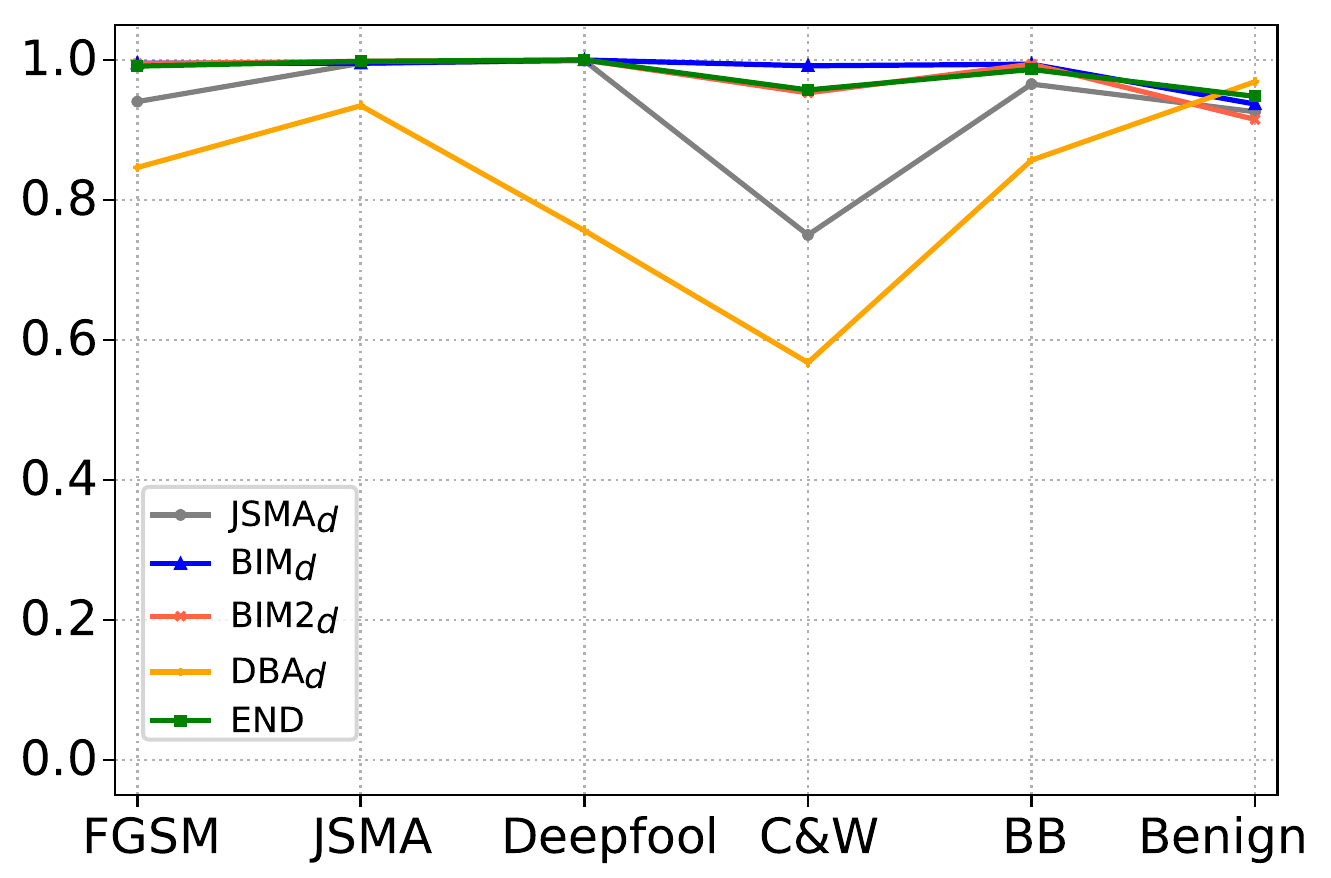}
}
\vspace*{-3mm}
\subfigure[K-NN, CIFAR10]{\label{fig:detectorknncifar10}
\includegraphics[width=0.225\textwidth]{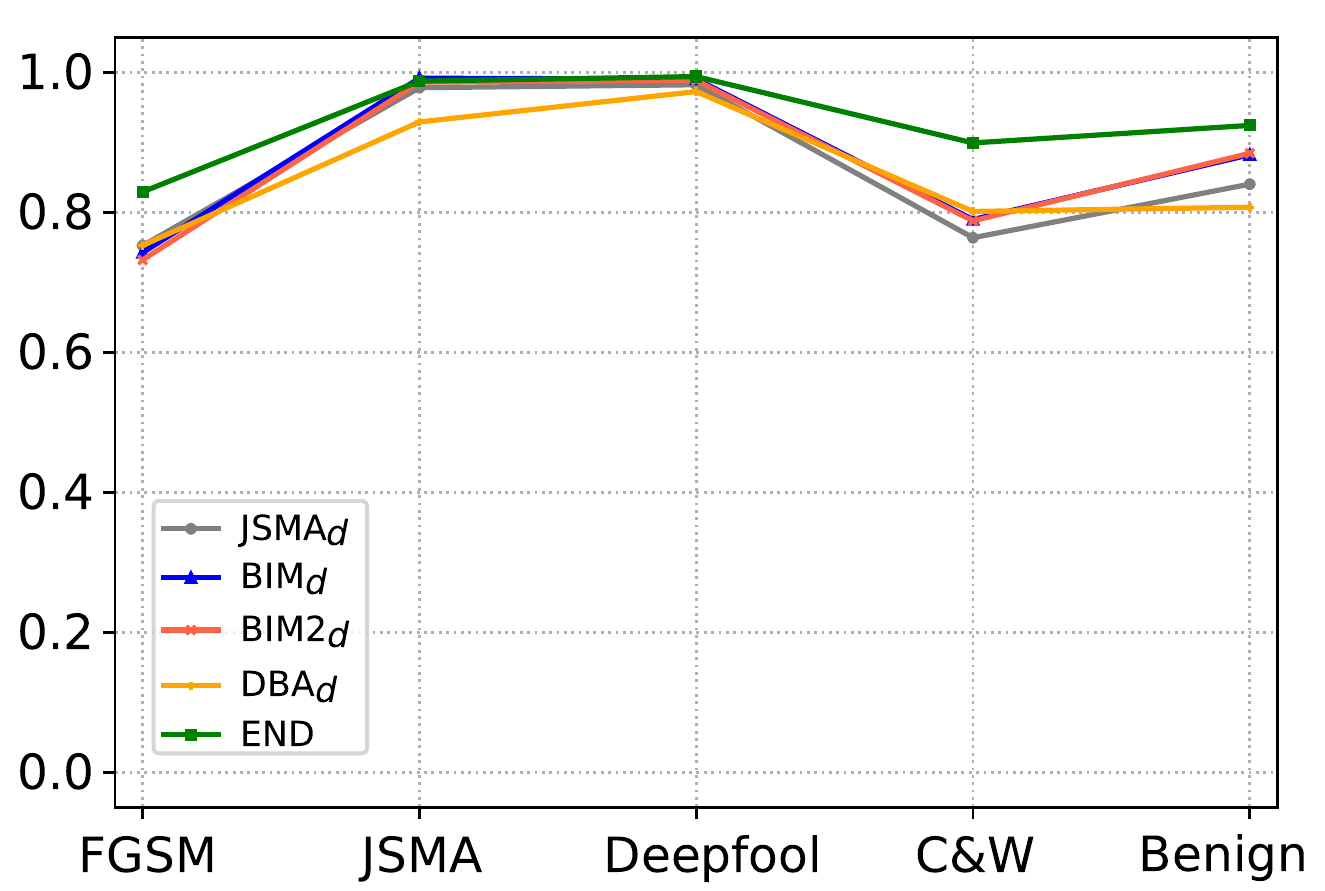}
}
\hspace*{-2mm}
\subfigure[Z-score, CIFAR10]{\label{fig:detectorzscorecifar10}
\includegraphics[width=0.225\textwidth]{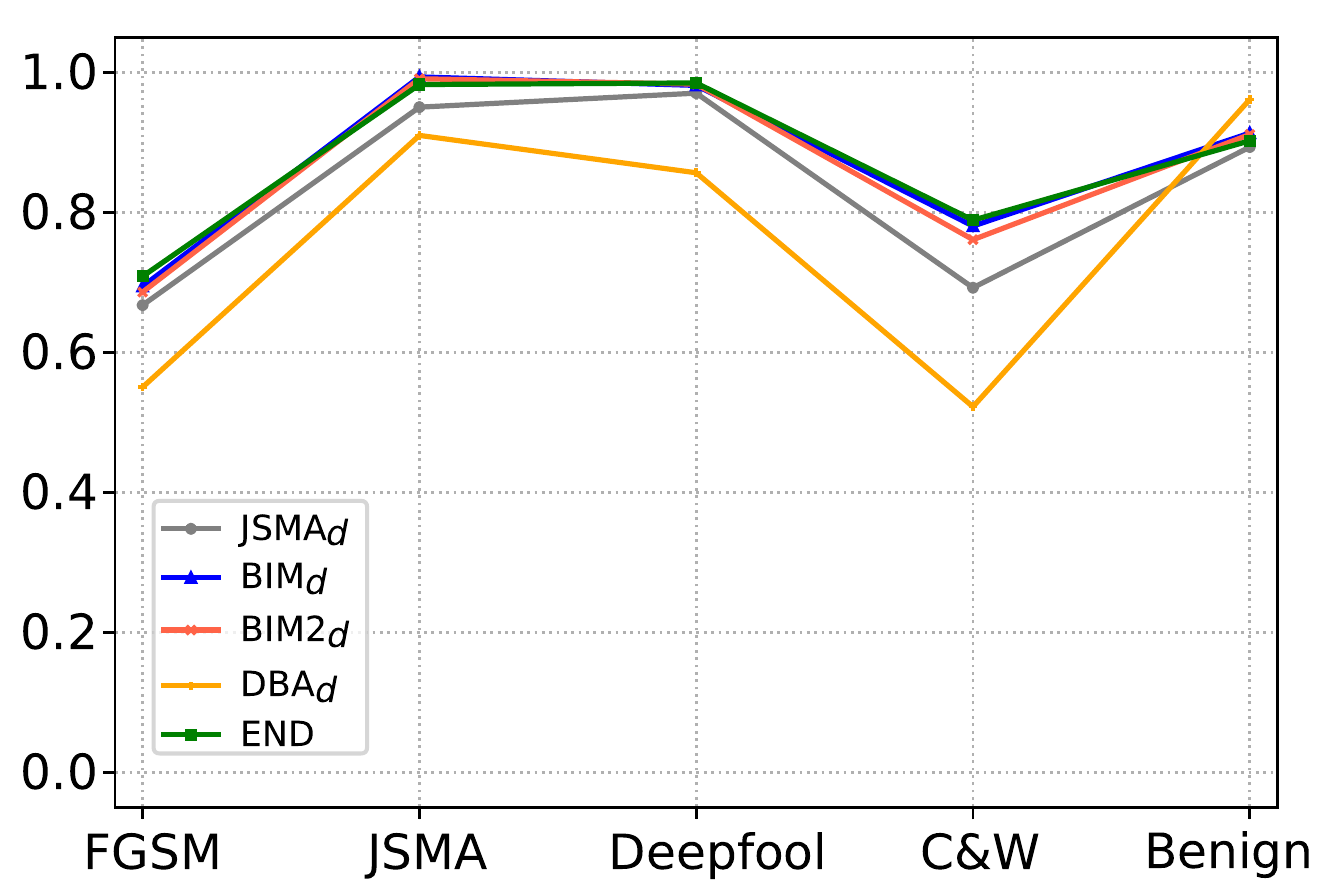}
}
\vspace*{-3mm}
\subfigure[K-NN, ImageNet]{\label{fig:detectorknnimagenet}
\includegraphics[width=0.225\textwidth]{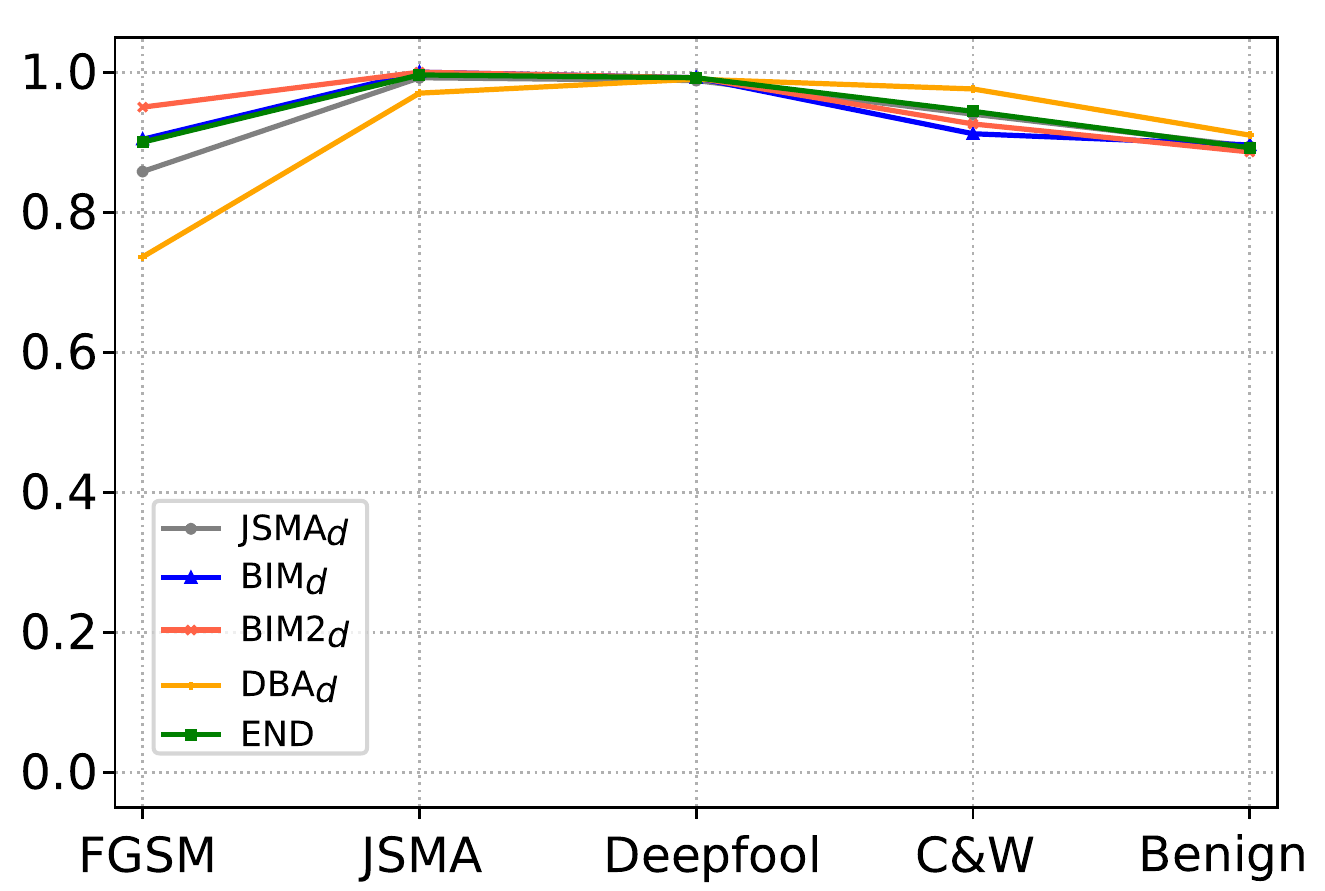}
}
\hspace*{-2mm}
\subfigure[Z-score, ImageNet]{\label{fig:detectorzscoreimagenet}
\includegraphics[width=0.225\textwidth]{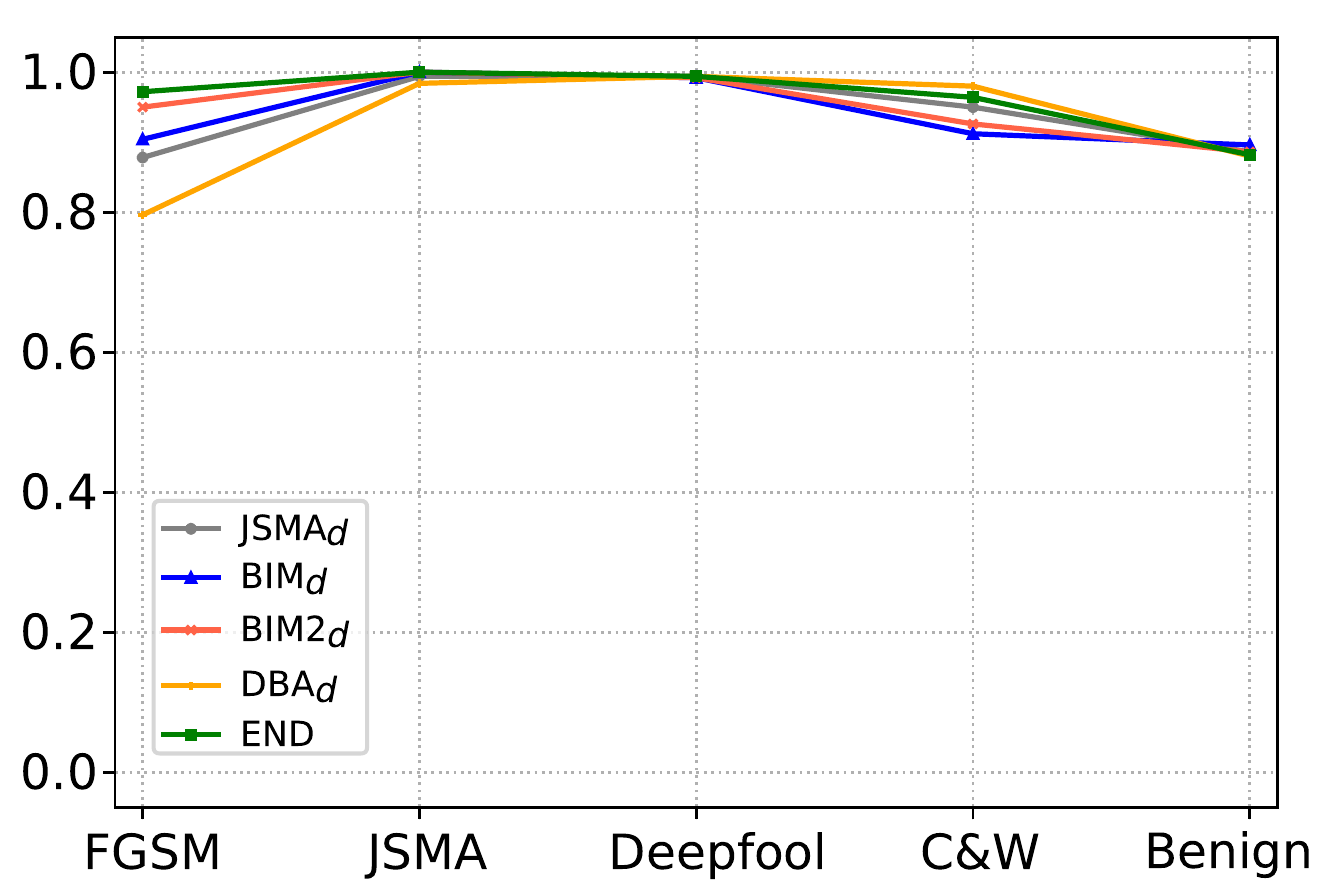}
}
\caption{Accuracy of detectors on different inputs}\vspace*{-4mm}
\label{fig:detector}
\end{figure}

The results are shown in Figures~\ref{fig:detectorknnmnist}, \ref{fig:detectorknncifar10} and \ref{fig:detectorknnimagenet}.
In general, on average, the accuracies of detectors JSMA$_{d}$, BIM$_{d}$, BIM2$_{d}$, DBA$_{d}$ and END are:
90.84\%, 98.09\%, 96.17\%, 87.42\% and 99.36\% for MNIST,
86.31\%, 87.90\%, 87.55\%, 85.23\% and 92.31\% for CIFAR10,
93.44\%, 94.08\%, 95.08\%, 91.64\% and 94.48\% for ImageNet,
We find that DBA$_d$ performs worse than others on most cases, which is consistent with
AUROC (cf. Table~\ref{table:auroc-17}).
It is worth noting that
though the ensemble detector END
does not always achieve the best performance,
it has the highest average accuracy. 
Thus, it balances the performances of individual detectors and is more robust.


\smallskip
\noindent
{\bf Z-score based detectors.}
For each dataset, and each attack$_d$ of BIM$_d$, BIM2$_d$, JSMA$_d$ and DBA$_d$ as defense,
we construct a Z-score based detector using the normal distribution of attack costs of 1,000 benign examples
via attack$_d$, resulting in detectors BIM$_{d}$, BIM2$_{d}$, JSMA$_{d}$ and DBA$_{d}$.
The threshold $h$ is -1.281552, which yields 10\% false positive rate on the 1,000 benign examples.
The ensemble detector named by END consists of these four detectors.
It classifies an input as benign if no less than 2 detectors classify the input as benign,
otherwise adversarial, namely, $k=2$. Results on tuning
$k$ and $h$ are given in Section~\ref{sec:parametertuning} in the supplementary material.

The results are shown in Figures~\ref{fig:detectorzscoremnist}, \ref{fig:detectorzscorecifar10} and \ref{fig:detectorzscoreimagenet}.
We can observe that they are able to achieve comparable or even better accuracy than K-NN based detectors,
although Z-score based detectors only use benign examples, whereas K-NN based detectors use
both benign and adversarial examples.


\subsubsection{Efficiency}
For a fair comparison with BL$_3$,
we report the detection costs of the Z-score based detectors here,
although it may be slightly worse than K-NN based detector.
The reason is that the threshold of Z-score detectors can be easily adjusted
to ensure that detection accuracy on benign examples is close to the baseline.
As both our method and the baseline BL$_3$ are query-intensive,
we compare the number of queries for efficiency comparison.


The results are reported in Table~\ref{table:time}.
Columns $\#_{\text{adv}}$ and $\#_{\text{benign}}$ give the number of queries to the model for adversarial and benign examples on average.
Columns Acc$_{\text{adv}}$ and Acc$_{\text{benign}}$ respectively give the accuracy for adversarial and benign examples on average.

By limiting the accuracy on benign examples to the one of BL$_3$,
we observe that all the white-box defenses (i.e., JSMA$_d$, BIM$_d$ and BIM2$_d$) outperform BL$_3$
in terms of the number of queries on both MNIST and CIFAR10.
Furthermore, they also achieve better accuracy than  BL$_3$ on CIFAR10,
while both BIM$_d$ and BIM2$_d$ achieve better accuracy than BL$_3$ on MNIST.
We also provide the results on ImageNet in Table~\ref{table:time}.
The results show that on ImageNet, \tool can still detect adversarial examples effectively and efficiently.
BIM$_d$/BIM2$_d$ are able to detect adversarial examples using one query.
This demonstrates that our method is also efficient on high-resolution images.
We finally remark that BL$_3$ does not support ImageNet
and the other baselines either provide only AUROC without constructing a detector
or do not provide cost analysis.

It is not surprising that the cost of DBA$_d$ is the largest one, as it is
a label-only black-box attack.
It is important to mention that black-box attacks used as defense
do not need any information of the models, hence
using black-box attacks as defense preserve the privacy of the models
while its effectiveness is still acceptable.

We emphasize that there is still space for optimization.
One latent optimization is to add an upper bound on the number of iterations, as adversarial examples often need fewer iterations than benign.
If the number of iterations reaches the bound and the attack fails,
the input can be considered as benign.
This optimization can reduce the number of iterations (hence queries) for benign examples
and both true and false positive rates will not be affected.

\begin{table}[t]
\centering
	\caption{Cost analysis of our detector with accuracy}
	\vspace*{-3mm}
	\label{table:time}
	\resizebox{0.42\textwidth}{!}{
		\begin{tabular}{c|c|c|c|c|c|c}
			\hline
			Dataset   & Detector & $\#_{\text{adv}}$ & Acc$_{\text{adv}}$    & $\#_{\text{benign}}$ &
			Threshold &
			Acc$_{\text{benign}}$ \\ \hline
			\multirow{5}{*}{MNIST}
			& BL$_3$    & 66 & 96.4\% & 463 & - & 89.7\%  \\
			\cline{2-7}
			& JSMA$_d$ & 20 & 95.4\% & 240 & 53 & \multirow{4}{*}{$\geq$ 89.7\%}  \\
			\cline{2-6}
			& BIM$_d$   & 16 & \textbf{99.8\%} & 148 & 122   \\
			\cline{2-6}
			& BIM2$_d$  & 38 & 99.6\% & 352 & 189    \\
			\cline{2-6}
			& DBA$_d$ & 92 & 88.2\% & 319 & 195   \\
			\hline
			\multirow{5}{*}{CIFAR10}
			& BL$_3$  & 67 & 90.6\% & 376 & - & 74.0\%  \\
			\cline{2-7}
			& JSMA$_d$ & 6 & 92.6\% & 33 & 13 & \multirow{4}{*}{$\geq$ 74.0\%} \\
			\cline{2-6}
			& BIM$_d$   & 14 & \textbf{93.0\%} & 65 & 35   \\
			\cline{2-6}
			& BIM2$_d$  & 29 & 92.7\% & 129 & 71   \\
			\cline{2-6}
			& DBA$_d$   & 252 & 87.7\% & 744 & 409  \\
			\hline
			\multirow{5}{*}{ImageNet}
			& BL$_3$    & - & - & - & - & -  \\
			\cline{2-7}
			& JSMA$_d$  & 6 & 95.4\% & 67 & 11 & 88.6\% \\
			\cline{2-7}
			& BIM$_d$   & 1 & 95.2\% & 4 & 2 & 89.6\%  \\
			\cline{2-7}
			& BIM2$_d$  & 1 & 96.7\% & 7 & 2 & 88.5\%  \\
			\cline{2-7}
			& DBA$_d$   & 143 & 93.9\% & 451 & 219 & 88.0\% \\
			\hline
		\end{tabular}
	}
	\vspace*{-3mm}
\end{table}

\smallskip
\noindent \textbf{Discussion}.
Here we briefly discuss how our approach can be used in practice
as different detectors have different accuracies.
Considering the tradeoff between the efficiency and accuracy,
one can use JSMA$_d$, BIM$_d$ or BIM2$_d$ as defense according to the dimension of images.
If one expects a more reliable and higher accurate detector,
an ensemble detector such as the END detector can be used.
If the privacy of the model matters,
a black-box attack based detector such as DBA$_d$ is better.

\begin{tcolorbox}[left=0mm,right=0mm,top=0mm,bottom=0mm]
\textbf{Answer to RQ3:} \tool is able to efficiently and effectively detect adversarial examples with lower false positive rate.
It is considerably more effective and efficient than the baseline BL$_3$.
\end{tcolorbox}

\subsection{Threats to Validity}
The threats to validity of our study include external and internal threats.
%
%
The selection of the subject datasets and target models could be one of the external threats. 
We tried to counter this issue by using 3 widely-used datasets and 5 pre-trained models
from well-established works.

The second external threat is the works we chose for comparison.
To mitigate this threat,
we compare with the works from both the artificial intelligence community
(e.g., BL$_1$ and BL$_2$)
and the software engineering community
(e.g., BL$_3$ and BL$_4$).
BL$_1$ is the most effective defense on MNIST among ten detections in~\cite{CW17a},
while BL$_2$ is claimed better than BL$_1$~\cite{LID}.
Both BL$_1$ and BL$_2$ are widely-used for comparison 
in the literature~\cite{shan2020gotta, ma2019nic, HeWCCS17, lee2018simple}. 
BL$_3$ and BL$_4$ are state-of-the-art methods from the perspective of software engineering.
It is worth noting that the comparison of baselines was conducted on the
repositories and parameters provided by the original authors,
to reproduce their best performance,
although it may be unfair to our method.

A further external threat is the knowledge of the adversary.
The same to baselines, we evaluated our approach against the original models
and assume that the adversary is unaware of the existence of the detection.
In practice, the adversary may learn that \tool has been used via social engineering or other methods,
and use a more threatening, specified attack method,
called adaptive attacks in~\cite{TCBM20}. We discuss and evaluate adaptive attacks against our defense method in Section~\ref{sec:adaptiveattacks}.
We do not perform the same adaptive attack on the baselines,
as adaptive attacks are usually designed for each specific defense method.

The internal threat mainly comes from the selection of attacks as defense.
We approximate model robustness of examples by the cost of attacking them
while attacks may differ in their capability.
To mitigate this threat,
we studied various attacks,
covering white-box and black-box attacks,
and $L_0$, $L_2$ and $L_\infty$ norm based attacks.
Experimental results indicate that our defense performs well regardless of the selected attacks,
although a minor difference can be observed.
Based on this, we conclude that our answers to research questions should generally hold.

\section{On Adaptive Attacks}
\label{sec:adaptiveattacks}
Lots of effective defenses have been shown to be ineffective in the presence of adaptive attacks~\cite{CW,CW17a,carlini2017magnet,HeWCCS17,TCBM20}.
Thus, adaptive attacks are the main threat to defense approaches.
In this section, we study possible adaptive attacks to our defense.



\subsection{Potential Bypass Approaches}
\subsubsection{Increasing Attack Costs}
A straightforward approach that may be used to bypass our defense is to increase the attack costs
so that the attack costs of adversarial and benign examples are similar.
To increase attack costs of adversarial examples, one can incorporate
attack costs into the loss function used to identify adversarial examples.
For instance, the adversary could change the loss function to
\begin{center}
$J'(x) = J(x) + \beta\cdot\max({\tt cost} - {\tt attack\_cost}(x), 0)$
\end{center}
where $J(x)$ is the original loss function,
$\beta$ is a parameter for balancing two terms of $J'(x)$,
${\tt cost}$ is the expected attack cost such as the mean of attack costs of benign examples or even the threshold of our Z-score based detection approach,
and ${\tt attack\_cost}(x)$ denotes the attack cost of $x$ via some attacks.
Minimizing the new loss increases the attack cost
until exceeding ${\tt cost}$.

\smallskip
\noindent \textbf{Discussion}.
This adaptive attack to our defense
is infeasible if not impossible.
First, the function ${\tt attack\_cost}(x)$
non-differentiable, consequently,
the loss function $J'(x)$ cannot be solved via gradient-based algorithms.
Second, non-gradient-based iterative attacks have to run some attacks
internally during each iteration in order to check if the attack succeeds or not.
This definitely results in high computational complexity, thus becoming infeasible .

\subsubsection{Increasing Robustness}
An alternative approach that may be used to bypass our defense is to
increase the robustness of adversarial examples, aiming to indirectly increase the attack costs.
However, it is non-trivial to directly control the robustness of adversarial examples.
%
We propose to increase the confidence/strength of adversarial examples,
initially considered by Carlini and Wagner~\cite{CW} for increasing transferability of adversarial examples
between different models.
Confidence is controlled by introducing a parameter
$\kappa$ into the loss function $J(x)$,
thus, the loss function becomes
\begin{center}
$J^\kappa(x) = \max(J(x),-\kappa)$
\end{center}
where the larger the parameter $\kappa$, the higher the confidence of the adversarial example.

The relation between robustness and confidence of adversarial examples is confirmed
by the following experiment.
We mount the C\&W attack on the previous 100 MNIST and 100 CIFAR10 images
under the same setting as~\cite{carlini2017magnet}, by varying the value of $\kappa$
and measuring the robustness using the CLEVER scores.
The results are reported in Table~\ref{table:tuningkappa}.
The experiment results show that
the adversary is able to increase the robustness of adversarial examples by increasing
the confidence. Therefore, high-confidence adversarial examples have the potential to bypass
our defense.

\begin{table}[t]
\centering
\caption{Robustness vs. confidence of adversarial examples}
\vspace*{-3mm}
\label{table:tuningkappa}
\resizebox{0.44\textwidth}{!}{\begin{tabular}{c|c|c|c|c|c|c}
  \hline
 \multicolumn{2}{c|}{ $\kappa$ } & 0 & 2 & 4 & 6 & 8 \\ \hline
\multirow{3}{*}{MNIST} & CLEVER Score   & $\approx 0$  & 0.11   &  0.14  &  0.14  & 0.17  \\  \cline{2-7}
& No. of Iterations  & 1.01  & 10.36   & 20.28  & 31.29  & 42.59  \\  \cline{2-7}
& $L_2$ distance  & 1.71  & 1.91   & 2.11   & 2.32   & 2.53  \\ \hline
  \multirow{3}{*}{CIFAR10} & CLEVER Score   & $\approx 0$  & 0.07   & 0.08   & 0.09   & 0.13  \\ \cline{2-7}
& No. of Iterations  & 1.37  & 8.53   & 17.29   & 24.47   & 34.4  \\  \cline{2-7}
& $L_2$ distance  & 0.41  & 0.52    & 0.67   & 0.82   & 0.99  \\
\hline
\end{tabular}}
\vspace*{-3mm}
\end{table}

\smallskip
\noindent \textbf{Discussion}.
This adaptive attack is feasible, but will introduce large distortion into adversarial examples when $\kappa$
increases, observed from Table~\ref{table:tuningkappa}.
As our defense changes neither inputs nor models,
it can be seamlessly combined with other defenses to defend against this adaptive attack.
\begin{itemize}[leftmargin=*]
    \item The first method is to combine with other defenses that are aimed at detecting adversarial examples with large distortion (e.g.,~\cite{MC17,RothKH19,LuIF17}).
    This would be able to detect a wide spectrum of adversarial examples.

    \item The second method is to combine with adversarial training~\cite{Intriguing,FGSM15,MyMSTV17} which enhances the DL model. Indeed, a successful attack to an adversarially trained model often introduces large distortion while the adaptive attack also introduces large distortion to bypass our defense, consequently,
        the distortion becomes too large to be human-perceptible.
\end{itemize}

\subsection{Evaluation of Adaptive Attacks}
Since the first adaptive attack is infeasible,
we only evaluate the second one which is implemented based on C\&W~\cite{carlini2017magnet}.
We evaluate this adaptive attack by varying
the parameter $\kappa$ from 0 to 20.

To evaluate the effectiveness of our defense combined with other defenses,
we consider the autoencoder based detector (AE for short)~\cite{MC17}
and PGD adversarial training (AT for short)~\cite{MyMSTV17}.
The AE trains a classifier $f_{ae}$ based on benign examples in order to
detect any adversarial examples with large distortion by checking
if $d(x,f_{ae}(x))$ is greater than a different threshold $\tau$,
where $d$ is a distance function, e.g., the mean squared error $\|x-f_{as}(x)\|_2$.



\begin{figure}[t]
\centering\vspace{-2mm}
\subfigure[\tool + AE]{
\includegraphics[width=0.235\textwidth]{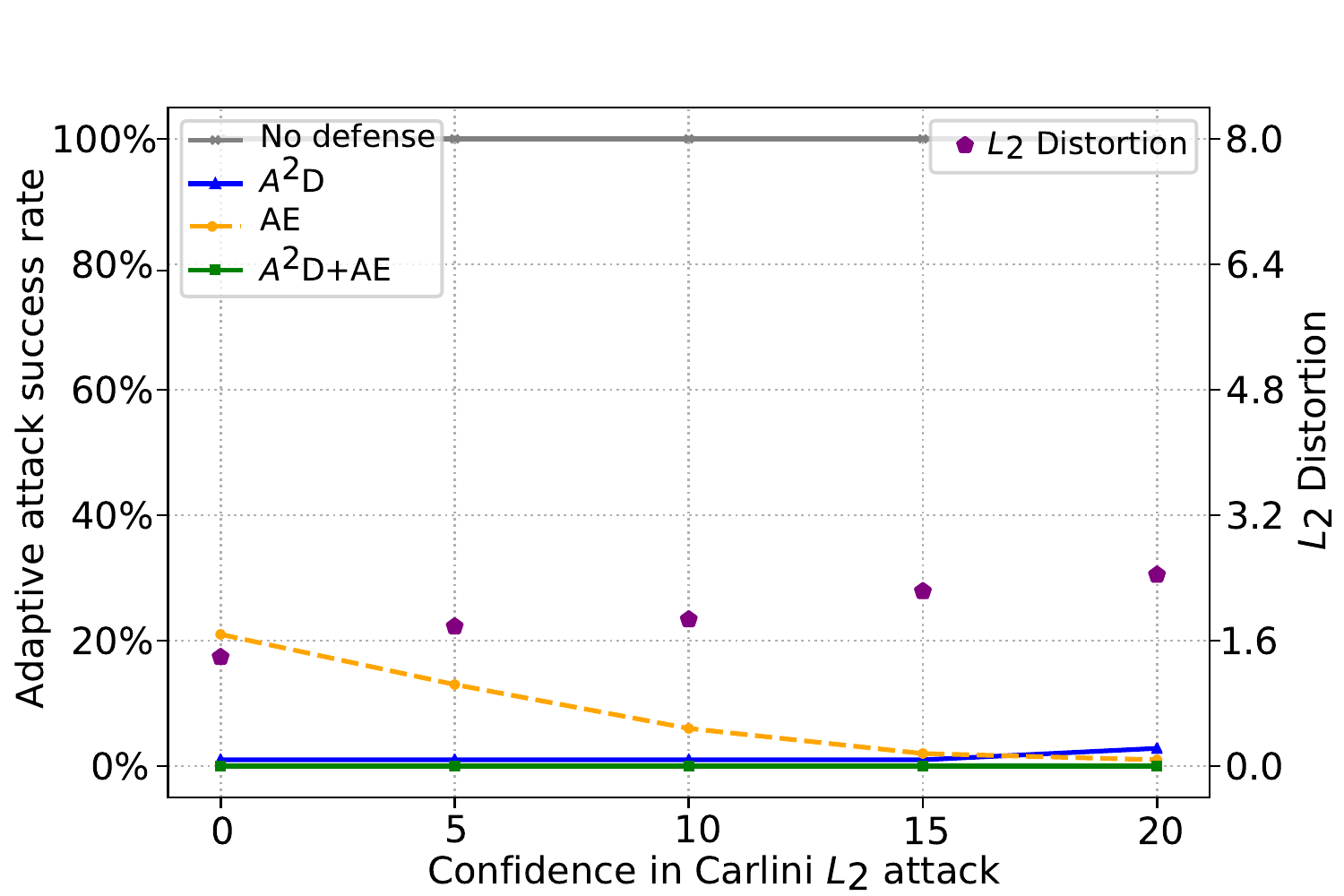}
\label{fig:aasd_AE}
}\hspace*{-1mm}
\subfigure[\tool + AT]{
\includegraphics[width=0.235\textwidth]{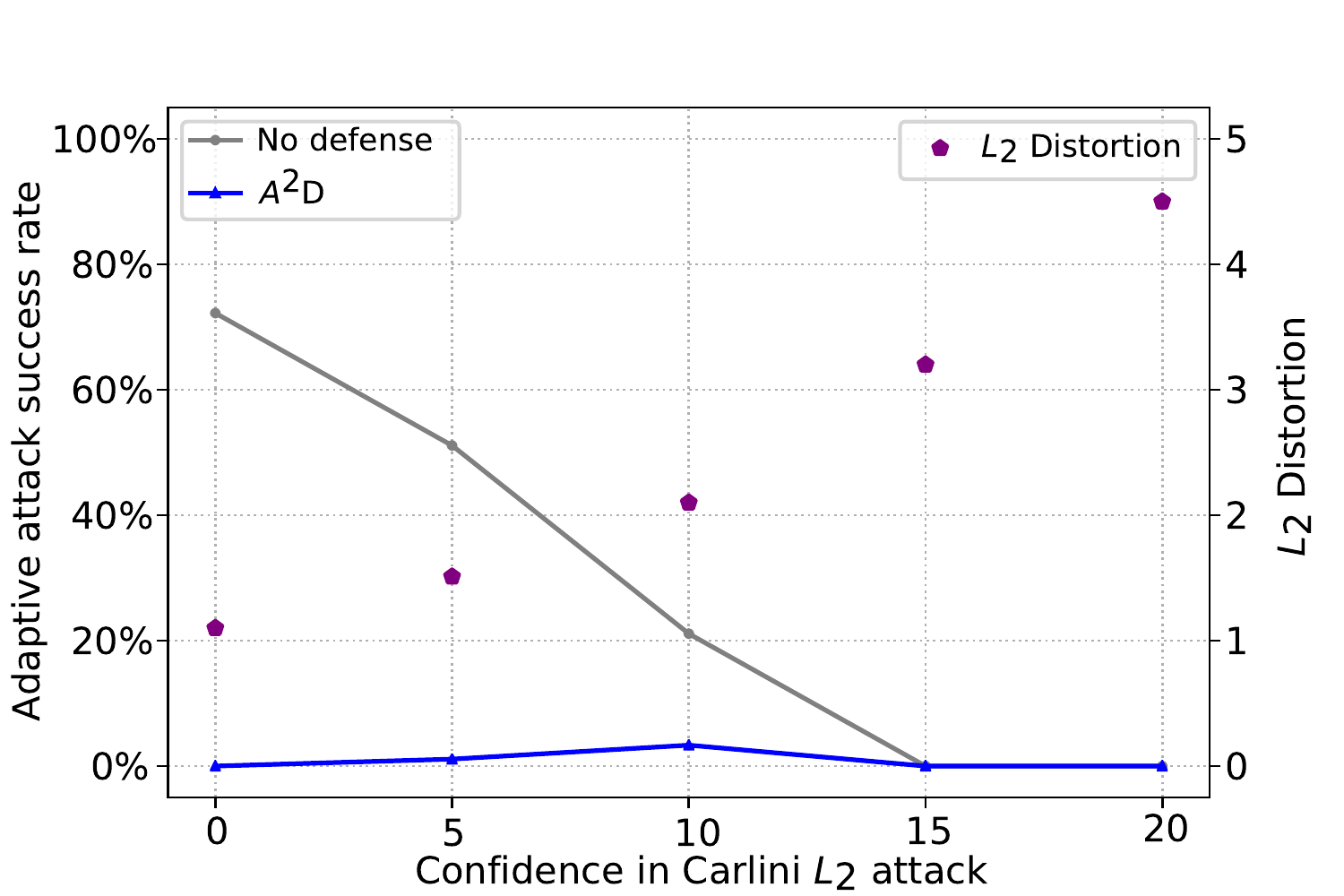}
\label{fig:aasd_AT}
}\vspace*{-4mm}
\caption{Adaptive attack results}
\label{fig:aasd_adaptive}\vspace*{-2mm}
\end{figure}

\subsubsection{\tool with AE}
\label{sec:a2d+ae}
For ease of evaluation, we conduct experiments
using the MNIST dataset under the same settings as~\cite{MC17} which provides a trained AE.
In our experiments, the maximal $L_2$ norm distortion
is 8.4 which is approximated from the maximal $L_\infty$ norm distortion
0.3 in Madry's challenges~\cite{mnistchallenge19}.
Note that our maximal $L_2$ distortion allows perturbations to be greater than the maximal $L_\infty$ distortion for some pixels.
Such large perturbations are often challenging for defense.
We use BIM$_d$ as defense
and the corresponding Z-score based detector
which only requires benign examples.
Thus, it is a relatively weaker defense.
We denote by \tool our detector
and \tool + AE the combined detector.

\smallskip
\noindent \textbf{Results.} The results are reported in Figure~\ref{fig:aasd_AE}.
From Figure~\ref{fig:aasd_AE} , we can observe that without any defense, the attack success rate (ASR) is always 100\%. 
With the increase of $\kappa$,
the detection rate of our defense \tool decreases slightly.
Specifically, \tool is able to detect all of the adversarial examples when $\kappa \leq 15$,
while only about 3\% of adversarial examples can bypass \tool when $\kappa=20$.
We also observe that both the $L_2$ distortion and detection rate of AE increase
with the increase of $\kappa$.
About 21\% of adversarial examples can bypass AE
when $\kappa=0$, while all adversarial examples can be detected by AE when $\kappa=20$.
Therefore, the ASR is always 0\% when the combined defense \tool + AE is applied.

\smallskip
\noindent \textbf{Summary.}
The above results demonstrate the benefit of combining two complementary defenses.
Although an adaptive attack can slightly reduce the effectiveness of \tool by increasing robustness of adversarial examples,
the combination of \tool and AE is able to completely defend against such adaptive attack.

\begin{figure}[t]
\centering
\includegraphics[width=0.45\textwidth]{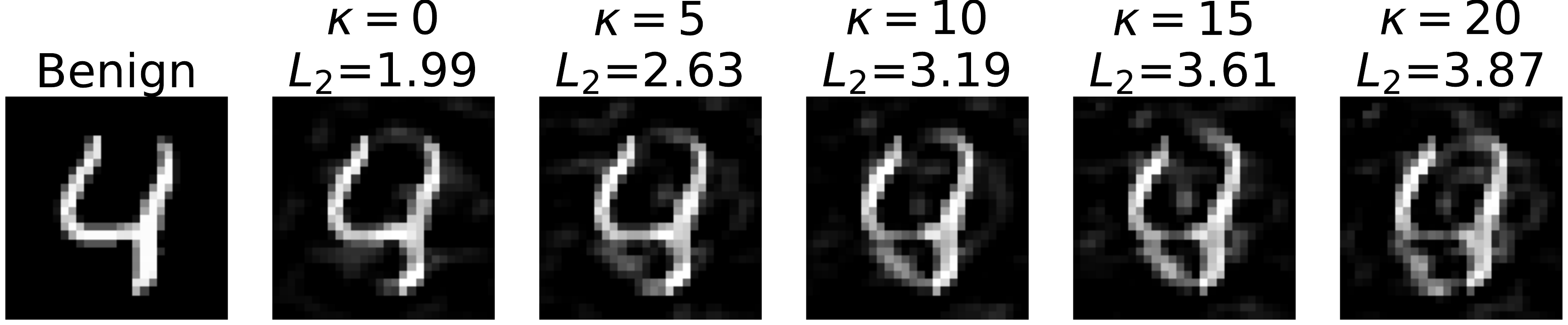}
\vspace{-3mm}
\caption{Adversarial examples for different $\kappa$}
\label{fig:mnist case}
\vspace{-3mm}
\end{figure}

\smallskip
\noindent \textbf{Case study.} Figure~\ref{fig:mnist case}
shows adversarial examples of a target attack from 4 to 0 with different $\kappa$.
The perturbation for $\kappa=20$ is about twice larger than
that for $\kappa=0$ and can be easily detected by AE.
More images are given in Section~\ref{sec:moreexamples} in the supplementary material.

\subsubsection{\tool with AT}
\label{sec:a2d+at}
For ease of evaluation,
we conduct experiments using the CIFAR10 dataset under the same settings as~\cite{MyMSTV17} which provides
an adversarially trained DL model.
In our experiments,
the maximal $L_2$ norm distortion is 1.6 which is approximated from the maximal $L_\infty$ norm distortion 0.03 in Madry's challenges~\cite{mnistchallenge19}.
We use the same Z-score based detector as in Section~\ref{sec:a2d+ae}.

\smallskip
\noindent \textbf{Results.}
The results are shown in Figure~\ref{fig:aasd_AT}.
Note that adversarial training (AT) is not a detector, so
`No defense' is equivalent to `AT'.
We can observe that AT is not very promising when
$\kappa$ is smaller, e.g., 72\% ASR for $\kappa=0$.
With the increase of $\kappa$ (i.e., increasing robustness of adversarial examples),
the ASR drops to 21\% when $\kappa=10$ and 0\% ASR when $\kappa \geq 15$.
This is because that finding adversarial examples
with distortion limited to the maximal $L_2$ threshold $1.6$
becomes more difficult for the adversarially trained model.
Recall that our defense \tool is good at detecting adversarial examples with small
distortion (i.e., low-confidence).
Therefore, the combined defense is very effective.
For instance, all adversarial examples with $\kappa=0$ can be detected by \tool, hence
the ASR drops from 72\% to 0\%.
The adaptive attack achieves no more than 3\% ASR on the adversarially trained model.

\smallskip
\noindent \textbf{Summary.}
To bypass our defense on adversarially trained models,
the adversary has to introduce much large distortion.
When perturbations are limited to human-imperceptible,
it becomes difficult to bypass our defense on adversarially trained models.

\begin{figure}[t]
\centering
\subfigure[Benign]{
\includegraphics[width=0.08\textwidth]{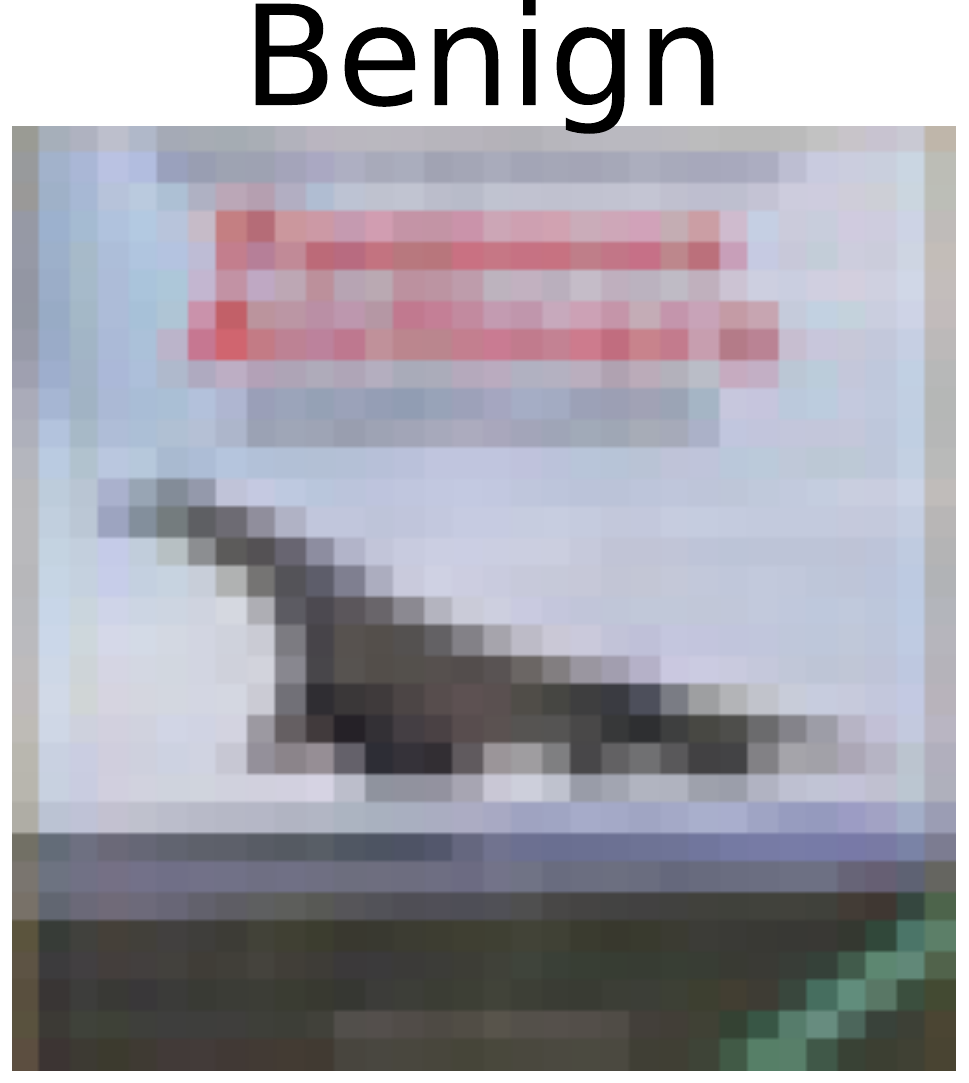}
}
\subfigure[]{
\includegraphics[width=0.08\textwidth]{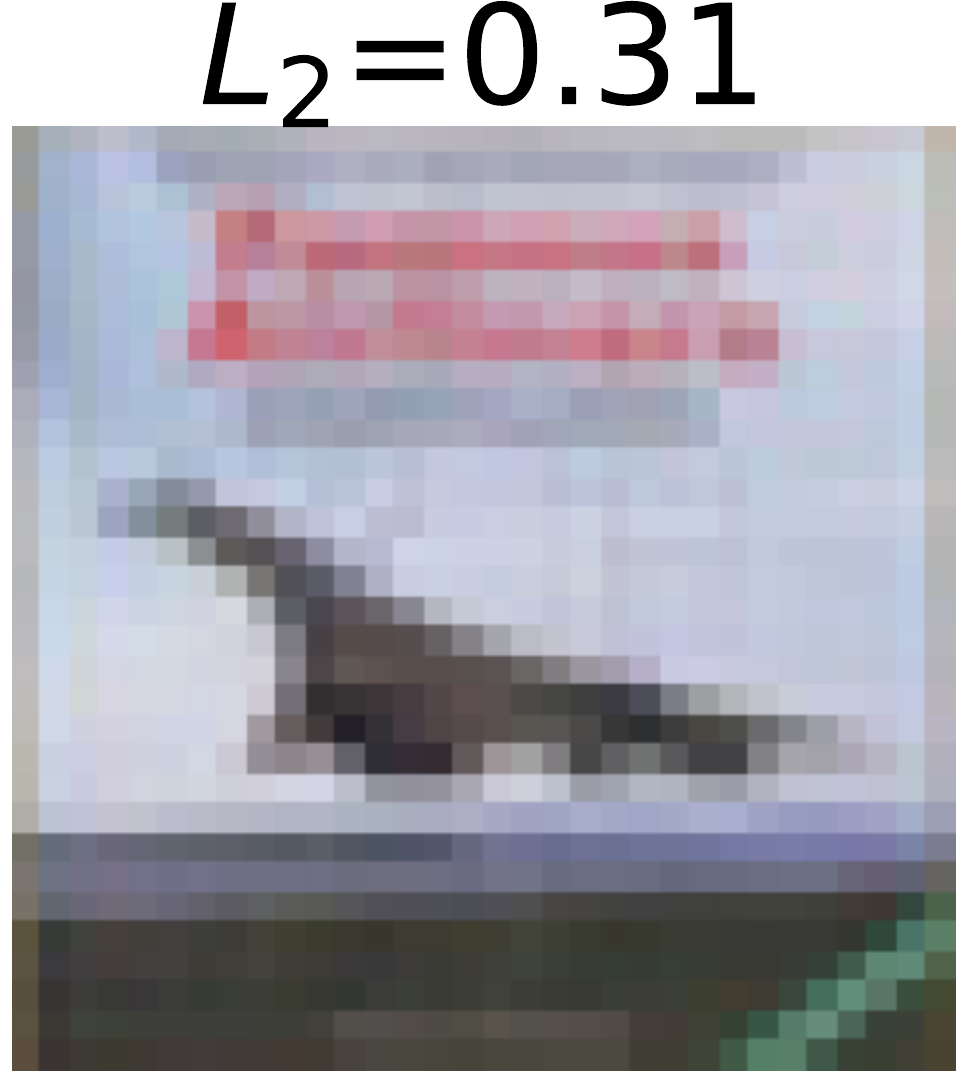}
}\subfigure[]{
\includegraphics[width=0.08\textwidth]{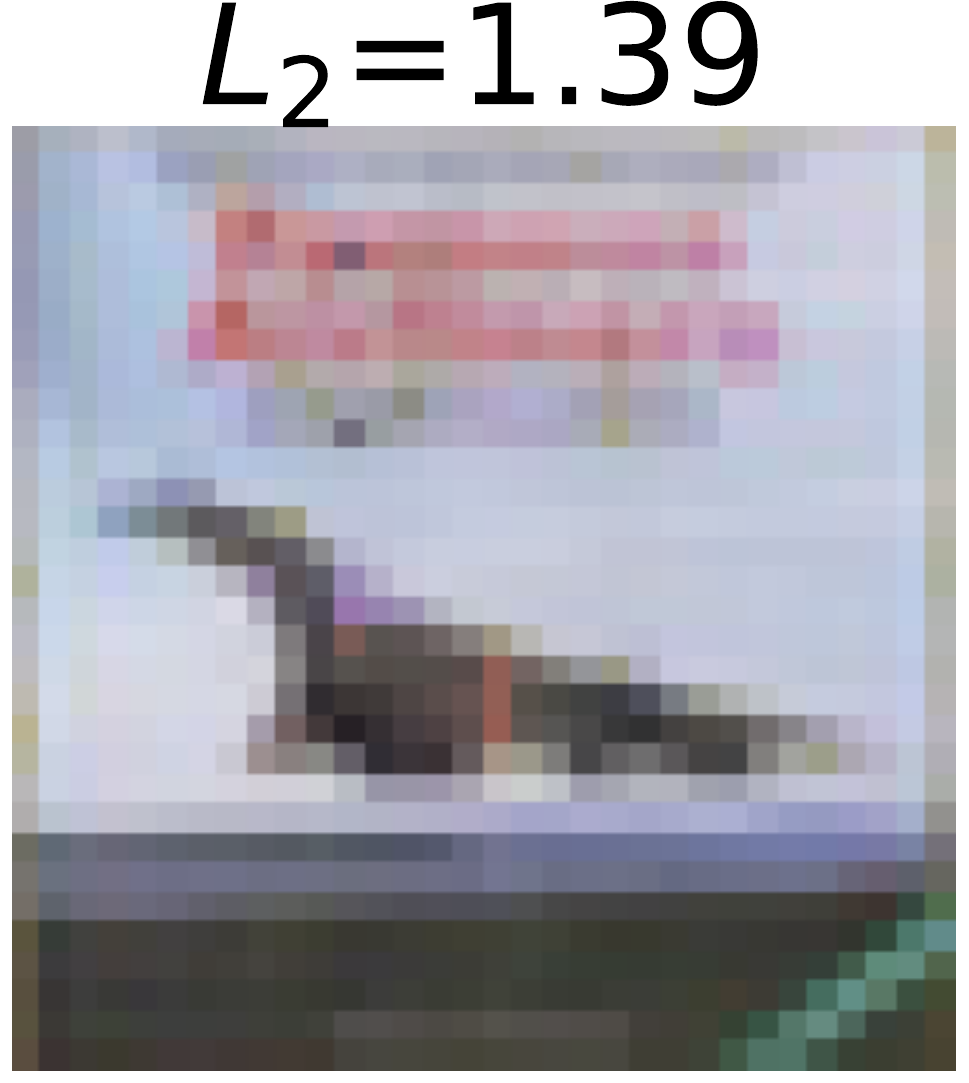}
}\subfigure[]{
\includegraphics[width=0.08\textwidth]{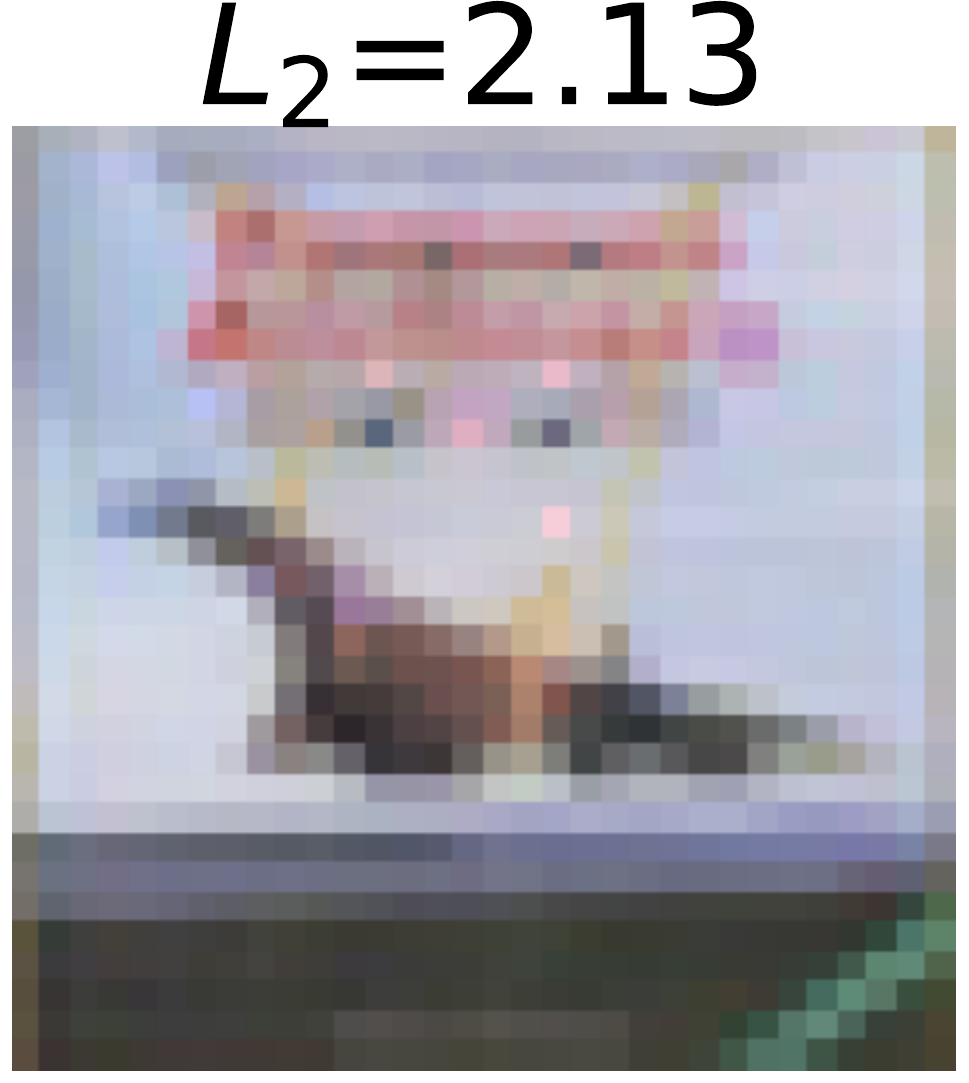}
}
\vspace{-3mm}
\caption{Adversarial examples
on the models without AT (b), with AT (c) and with \tool + AT (d)}
\label{fig:target attack case}
\vspace{-3mm}
\end{figure}

\smallskip
\noindent \textbf{Case study.}
Figure~\ref{fig:target attack case} shows
adversarial examples of a target attack from
`airplane' to `cat'.
Without any defense,
an adversarial example with less distortion can be crafted, cf. Figure~\ref{fig:target attack case}(b).
With AT, it requires more distortion to craft an adversarial example, cf. Figure~\ref{fig:target attack case}(c).
If both \tool and AT are enabled,
it requires much more distortion to craft adversarial examples, cf. Figure~\ref{fig:target attack case}(d).
Now the distortion is too large to be human-perceptible,
and we can clearly see the silhouettes of `cats' on the adversarial example.
More images are given in Section~\ref{sec:moreexamples} in the supplementary material.

\section{Related Work}
\label{sec:related}
As a new type of software system,
neural networks have received extensive attention over the last five years.
We classify existing works along three dimensions:
adversarial attack, adversarial defense, and neural network testing \& verification.

\smallskip
\noindent {\bf Adversarial attack.}
Adversarial attacks aim to misjudge the neural network
by adding perturbations that are imperceptible to humans.
We have introduced common attacks in Section~\ref{sec:back}.
We selected multiple types of methods to generate adversarial examples,
FGSM~\cite{FGSM15}, BIM~\cite{KGB17},
JSMA~\cite{JSMA},
DeepFool~\cite{DeepFool}, C\&W~\cite{CW} and substitute model attack~\cite{PMG16}.
We also selected multiple attacks as defense.
Meanwhile, several adversarial examples have the ability to
carry out attacks in the real environment~\cite{KGB17, EEF0RXPKS18, JMLHW19},
and pose threats to neural networks, which is a new programming paradigm.

\smallskip
\noindent {\bf Adversarial defense.}
A typical defense method is adversarial training
~\cite{Intriguing,FGSM15,KGB17,MyMSTV17}, which
produces adversarial examples and injects them into training data.
Another type of defenses protects models by pre-processing the input data~\cite{buckman2018thermometer,guo2017countering, xie2017mitigating}
or projects potential adversarial examples onto the benign data manifold before classifying them~\cite{song2017pixeldefend, MC17}.
Detection is another defense approach to adversarial examples.
If an input is detected as adversarial, it will be rejected without being fed to the model~\cite{feinman2017detecting,lee2018simple,LID,Xu0Q18}.
However, it has been shown that most defense methods
except adversarial training can be easily bypassed by adaptive attack with
backward pass differentiable approximation
and expectation over transformation~\cite{athalye2018obfuscated,AEIK18,TCBM20,CW17a}.
Until today this problem is still of considerable interest.
However, most detection methods are not effective on high-resolution images
(e.g., ImageNet dataset)~\cite{wang2019adversarial, MC17, LID, feinman2017detecting} 
or do not consider adaptive attacks~\cite{wang2019adversarial, MC17, LID, feinman2017detecting, wang2019adversarial}.
Our defense method is effective on 3 widely-used datasets
covering both low- and high-resolution images. We also evaluated
our defense against potential adaptive attacks and demonstrate
its effectiveness.

\smallskip
\noindent {\bf Neural network testing \& verification.}
Some works look for vulnerabilities in neural networks from the perspective of software testing.
For example, DeepXplore~\cite{PCYJ17} proposed a testing technique to find adversarial examples guided neuron coverage.
After that, a series of coverage criteria have been proposed for neural network testing~\cite{SHK18, MJZSXLCSLLZW18, kim2019guiding}.
Other testing methods also have been adapted to test neural networks such as concolic testing~\cite{SWRHKK18}, mutation testing~\cite{MZSXLJXLLZW18, shen2018munn}, 
and so on~\cite{WHK18, MaJXLLLZ19}.
Some testing methods focus on different application scenarios of neural networks, including DeepTest~\cite{TPJR18},
DeepRoad~\cite{zhang2018deeproad},
Deepbillboard~\cite{zhou2020deepbillboard} and object-relevance metamorphic testing~\cite{tian2019testing}.
Some other works focus on testing the neural network at the architecture level~\cite{zhang2020detecting} or the deep learning library itself~\cite{wang2020deep}.
We do not use testing criteria to model the robustness of examples,
as testing criteria are not necessarily correlated with robustness~\cite{yan2020correlations,dongempirical}.

Various formal verification techniques have been proposed
to verify robustness property against neural networks~\cite{HKWW17, GKPB18, PT10, KBDJK17, wang2018efficient, BILVNC16, DSGMK18, TXT18, EGK20, SGPV19, SGMPV18, WZCSHDBD18, Ehl17, GMDTCV18}.
Formal verification provides provable or theoretic guarantees,
and robustness is also the source of our defense approach.
However, formal verification suffers from the scalability problem,
due to the high computational complexity. 
Therefore, we used attack cost instead.

\cite{wang2019adversarial} and~\cite{Wang2019Dissector} are very close to our defense approach.
They also considered the problem of
identifying adversarial examples from the perspective of software engineering,
by leveraging mutation testing and model anatomy respectively.
However, both of them have to modify the original model,
while our defense approach does not,
hence is easy to deploy.
When using white-box attacks as defense, 
the model only needs to provide an interface for logits and gradients,
rather than model parameters. When using black-box attacks as defense,
the model only needs to provide the classification results,
thus protecting the privacy of the model.
Inspired by~\cite{carlini2017magnet,HeWCCS17,TCBM20,ma2019nic},
we discussed and evaluated adaptive attacks to our defense.
However, \cite{wang2019adversarial} and~\cite{Wang2019Dissector}
do not consider adaptive attacks, hence it is unclear whether 
they are still effective under adaptive attacks.

\section{Conclusion}
We have proposed a novel characterization of adversarial examples via robustness.
Based on the characterization, we proposed a novel detection approach, named attack as defense (\tool), 
which utilizes existing attacks to measure examples' robustness.
We conducted extensive experiments to evaluate our observations and detection approach \tool, showing that it outperforms four recent promising approaches.
We also thoroughly discussed the main threat (i.e., adaptive attacks) to our defense and evaluated them to our defense.
By combing our defense with an existing defense and adversarial training,
the results are very promising, e.g., the ASR drops from
72\% to 0\% on CIFAR10,
and drops from 100\% to 0\% on MNIST.
\label{sec:concl}



\appendix

\section{Environments}
\label{sec:appendix-env}
For reproductivity of this work, 
all the information of the target models and attack parameters used in our experiments are given below.

For Env$_1$, we directly use models and attack methods provided by BL$_1$~\cite{feinman2017detecting},
the DL model for MNIST is LeNet,
the DL model for CIFAR10 is a deep 12-layer convnet.
The accuracy of the target model on training/testing dataset is 99.6\%/99.1\% for MNIST and 87.3\%/80.3\% for CIFAR10.
BL$_2$~\cite{LID} uses the models and attack code segments
provided by BL$_1$,
so Env$_1$ is the environment used by these two baselines.
It should be noted that there are two slightly different BIM implementations in Env$_1$, and we use the `bim-a'.

For Env$_2$,
we integrated the models and datasets used by BL$_3$~\cite{wang2019adversarial} and BL$_4$~\cite{Wang2019Dissector}.
These two papers were published in ICSE'19 and ICSE'20 respectively, and both use PyTorch as the platform.
BL$_3$ provides models of MNIST~(LeNet) and CIFAR10~(GooglLeNet),
the accuracy of the target model on training/testing dataset is 98.5\%/98.3\% for MNIST and 99.7\%/90.5\% for CIFAR10,
and BL$_4$ provides a model of ImageNet~(ResNet101),
which has a top-1 accuracy rate 77.36\% on the validation set.
In addition,
BL$_4$ only provides one attack that is $L_2$ norm adoption of FGSM with 0.016 as the attack step.
In order to make the results more comprehensive,
we used the three attacks JSMA, DeepFool and C\&W, implemented in Foolbox, to generate adversarial examples with default parameters.

\begin{table}[t]
\centering
\caption{Parameters of attacks for Env$_1$}
\vspace{-3mm}
\label{table:env1-attack}
\resizebox{0.34\textwidth}{!}{
\begin{tabular}{c|c|c}
\hline
Dataset   & Attack Method & Parameters    \\ \hline
\multirow{4}{*}{\tabincell{c}{MNIST}} & FGSM & $\epsilon=0.3$  \\
\cline{2-3}
& BIM   & $\epsilon=0.3, \alpha=0.01$    \\
\cline{2-3}
& JSMA  & $\theta=1, \gamma=0.1$  \\
\cline{2-3}
& C\&W    & $\kappa=0, c=0.02$
\\
\hline
\multirow{4}{*}{\tabincell{c}{CIFAR10}} & FGSM & $\epsilon=0.05$  \\
\cline{2-3}
& BIM       & $\epsilon=0.05, \alpha=0.005$    \\
\cline{2-3}
& JSMA  & $\theta=1, \gamma=0.1$    \\
\cline{2-3}
& C\&W    & $\kappa=0, c=0.02$   \\
\hline
\end{tabular}}
\end{table}

\begin{table}[t]
\centering
\caption{Parameters of attacks for Env$_2$}\vspace{-3mm}
\label{table:env2-attack}
\resizebox{0.37\textwidth}{!}{\begin{tabular}{c|c|c}
\hline
Dataset   & Attack Method & Parameters  \\ \hline
\multirow{5}{*}{\tabincell{c}{MNIST}} & FGSM & $\epsilon=0.35$ \\ \cline{2-3}
& JSMA   & $\theta=1, \gamma=0.12$  \\ \cline{2-3}
& DeepFool  & overshoot=0.02 \\ \cline{2-3}
& C\&W    & $\kappa=0, c=0.6$  \\ \cline{2-3}
& BB   & Sub model+FGSM, $\epsilon=0.35$ \\ \hline
\multirow{4}{*}{\tabincell{c}{CIFAR10}} & FGSM & $\epsilon=0.05$   \\\cline{2-3}
& JSMA       & $\theta=1, \gamma=0.12$     \\\cline{2-3}
& DeepFool  & overshoot=0.02 \\ \cline{2-3}
& C\&W  & $\kappa=0, c=0.6$   \\
\hline
\multirow{4}{*}{\tabincell{c}{ImageNet}}
& FGSM  & $\epsilon=0.016$, $L_2$ norm
\\\cline{2-3}
& JSMA  & \multirow{3}{*}{\tabincell{c}{Foolbox default parameters}}
\\\cline{2-2}
& DeepFool   &
\\ \cline{2-2}
& C\&W  &
\\
\hline
\end{tabular}}
\end{table}

\section{Attack Time vs. Number of Iterations}
\label{sec:itervstime}

\begin{figure}[t]
  \centering \vspace{-3mm}
    \subfigure[Time vs. $\sharp$iter on MNIST]{\label{fig:scorevsattackiterMNIST}
    \includegraphics[width=0.235\textwidth]{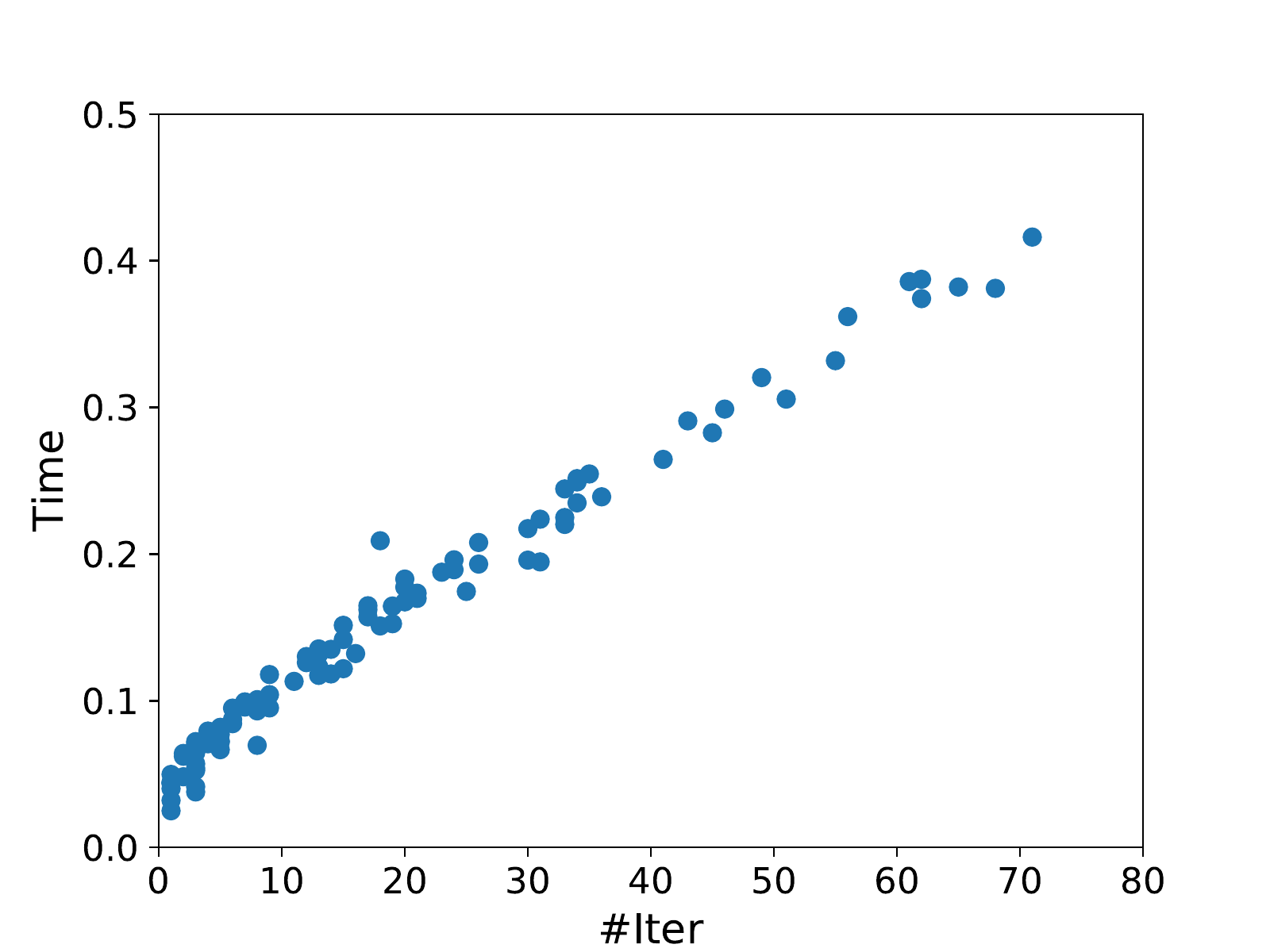}
    }\hspace*{-2mm}
  \subfigure[Time vs. $\sharp$iter on CIFAR10]{\label{fig:scorevsattackiterCIFAR}
    \includegraphics[width=0.235\textwidth]{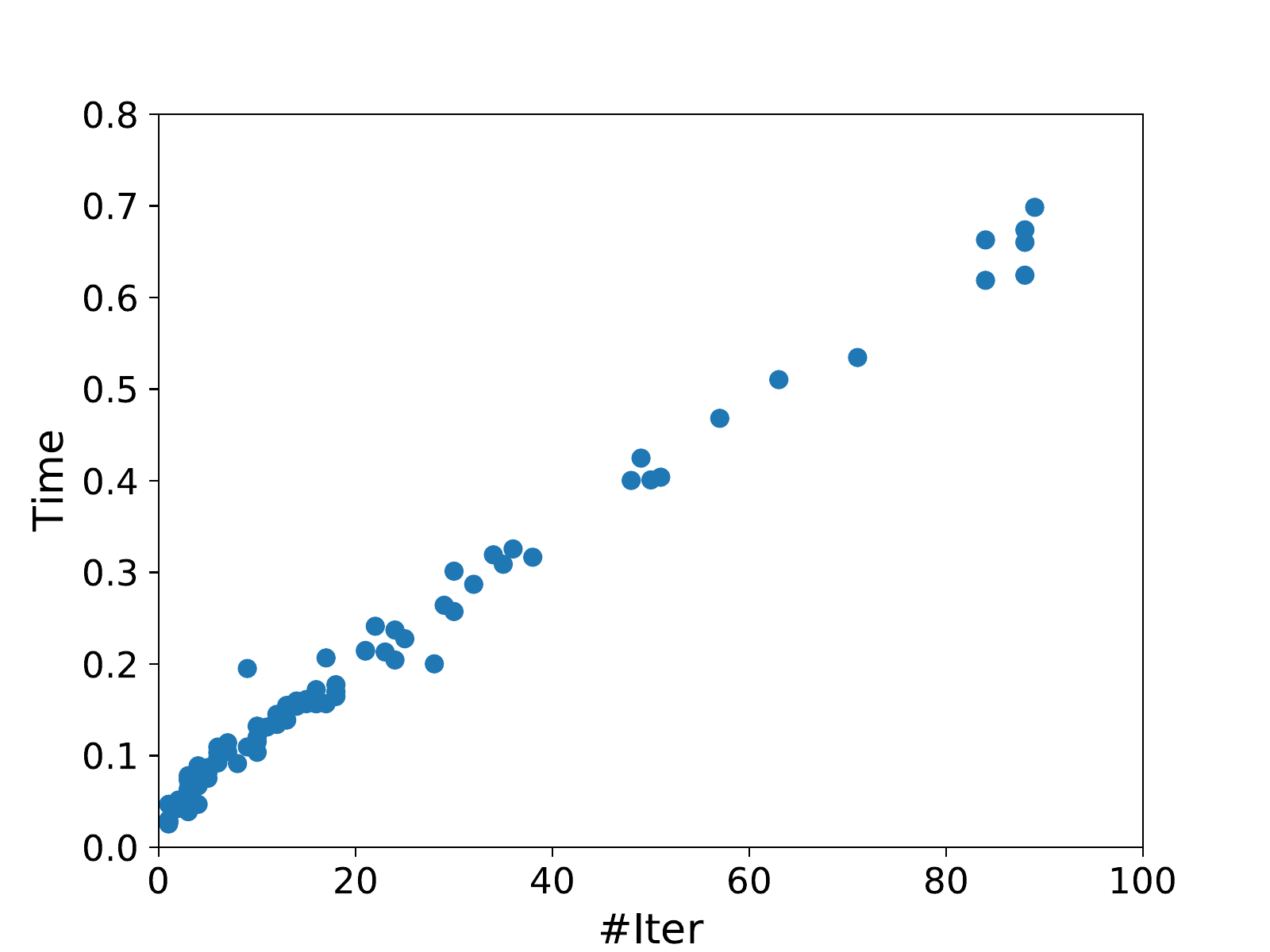}
    }\vspace*{-2mm}
  \caption{Attack time vs. iterations ($\sharp$iter) using JSMA}
\end{figure}

For an iterative attack, the number of iterations is positively correlated with attack time.
This is justified by the scatter plots shown in Figure~\ref{fig:scorevsattackiterMNIST} and Figure~\ref{fig:scorevsattackiterCIFAR},
which compare the attack time and number of iterations by JSMA
on the 100 MNIST and 100 CIFAR10 examples.

\section{Results of Parameter Tuning}
\label{sec:parametertuning}
We report results of turning parameters
on the target model from Env$_1$ using the MNIST images.


\begin{table}[t]
\centering
\caption{Comparison for the impact of classifier parameters}\vspace{-3mm}
\label{table:parameter}
 \resizebox{0.36\textwidth}{!}{
    \begin{tabular}{c|c|c|c}
    \hline
    Parameter   & Value    &  Acc$_{\text{benign}}$     &  Acc$_{\text{adv}}$    \\ \hline
    \multirow{7}{*}{K-value}
    & 5   & 0.988	&  	0.9373				
    \\ \cline{2-4}
    & 10     & 0.986  &  0.9465
    \\ \cline{2-4}
    & 25     & 0.986  & 0.9545
    \\ \cline{2-4}
    & 50     & 0.985  & 0.962   \\
    \cline{2-4}
    & \bf{100}   & \textbf{0.979}  & \textbf{0.9685}    \\
    \cline{2-4}
    & 150    & 0.977  & 0.9715    \\
    \cline{2-4}
    & 200    & 0.976  &  0.9753   \\
    \hline

    \multirow{5}{*}{\tabincell{c}{Ratio between\\ benign and\\ adversarial\\ examples}}
    & 1:0.5  & 0.981 & 	0.9643				
    \\ \cline{2-4}
    & 1:0.8  & 0.979 & 	0.9658		
    \\ \cline{2-4}
    & \bf{1:1}  & \textbf{0.979} & \textbf{0.9685}
    \\ \cline{2-4}
    & 1:1.2  &  0.977 & 0.9705
    \\ \cline{2-4}
    & 1:2 & 0.971 & 0.9768
    \\ \hline

    \multirow{5}{*}{\tabincell{c}{Z-Score}}
    & -1  & 0.875 & 0.9918			
    \\ \cline{2-4}
    & \bf{-1.281552}  & \textbf{0.926} & \textbf{0.987}
    \\ \cline{2-4}
    & -1.644854  & 0.963 & 0.9763
    \\ \cline{2-4}
    & -1.959964 & 0.979 & 0.9635
    \\ \cline{2-4}
    & -2  & 0.98 & 0.962
    \\ \hline

    \multirow{4}{*}{\tabincell{c}{Statistic\\ Ensemble}}
    & 1  & 0.972 & 0.8835			
    \\ \cline{2-4}
    & \bf{2}  & \textbf{0.926} & \textbf{0.987}
    \\ \cline{2-4}
    & 3  & 0.883 & 0.994
    \\ \cline{2-4}
    & 4 & 0.775 & 0.9963
    \\ \hline
    \multirow{4}{*}{Classifier}
    & {\bf K-NN}    & \textbf{0.979}    & \textbf{0.9685}	
    \\ \cline{2-4}
    & SVM    & 0.975    & 0.9643 \\
    \cline{2-4}
    & DTC    & 0.994    & 0.9078   \\
    \cline{2-4}
    & RFC    & 0.994    & 0.9068  \\
    \hline
    \end{tabular}
    }
\end{table}

We vary the value of $K$ from 5 to 200 for the K-NN based ensemble detector.
The results are shown in Table~\ref{table:parameter}, where the \textbf{bold} one denotes
the value used in the previous experiments.
 Column \emph{Acc$_{\text{adv}}$} denotes
the accuracy on adversarial examples. Column \emph{Acc$_{\text{benign}}$} denotes
the accuracy on benign examples.
We observe that both true and false positive rates slightly increase with the increase of $K$.
We also vary the ratio between benign and adversarial examples.
As shown in Table~\ref{table:parameter}, with the decrease of the ratio, both true and false positive rates slightly increase.

For the Z-score based ensemble detector, we vary the value of the threshold $h$
from -1, -1.281552, -1.644854 and -2, which respectively correspond to
about 15\%,  10\%, 5\% and 2.3\% false positive rates of 1,000 benign samples.
From Table~\ref{table:parameter}, we observe that
the smaller the threshold $h$, the lower the false positive rate, but the higher
the true positive rate.
We also vary the parameter $k$ in the Z-score based ensemble detector.
Recall that the ensemble detector classifies an input to benign if $k$ detectors classify the input to benign, otherwise adversarial.
We observe from Table~\ref{table:parameter} that
both true and false positive rates increase  with the increase of $k$.

In summary, the above parameters can be used to balance the true and false positive rates, namely,
the true positive rate could be improved at the cost of false positive rate.

Finally, instead of K-NN, we also tried the support vector machine (SVM), decision tree (DTC)
and random forest (RFC) classification algorithms.
We use the implementations of scikit-learn with the default parameters\footnote{https://scikit-learn.org}.
The results in Table~\ref{table:parameter} show that
similar accuracy can be obtained using different classification algorithms.
This implies that our detection approach is generic in terms of classification algorithms.

\section{More Examples}
\label{sec:moreexamples}

\begin{figure*}
    \centering
    \includegraphics[width=0.95\textwidth]{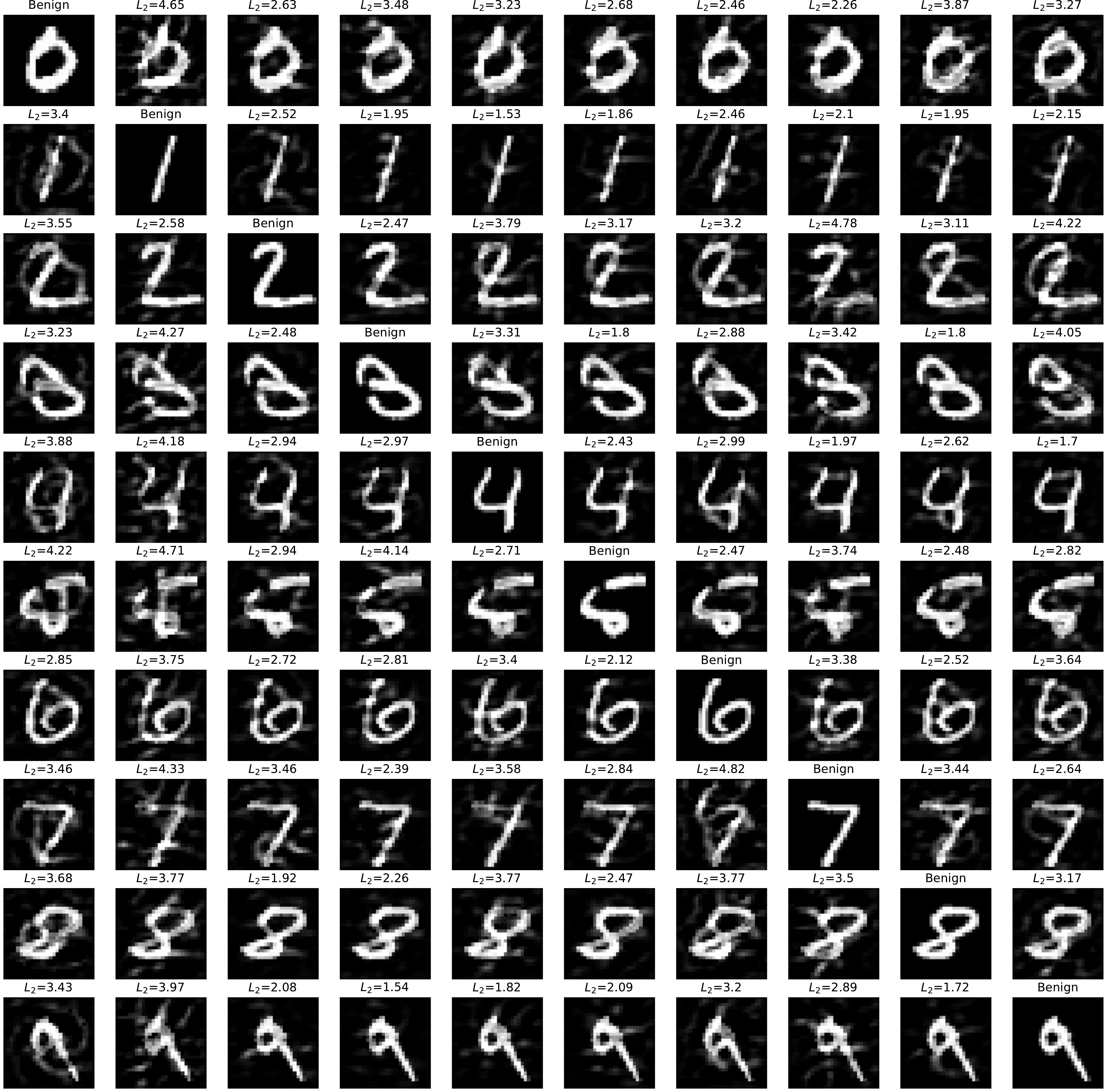}
    \caption{Targeted adversarial examples for each label pair of images on the MNIST model, $\kappa=20$}
    \label{fig:mnist_k20}
\end{figure*}

Figure~\ref{fig:mnist_k20} shows adversarial examples under the attack in Section~\ref{sec:a2d+ae} with
$\kappa=20$, where each of them except for the benign one targets
one of the other labels.
All these adversarial examples are successfully detected by \tool + AE.

\begin{figure*}
    \centering
    \includegraphics[width=0.95\textwidth]{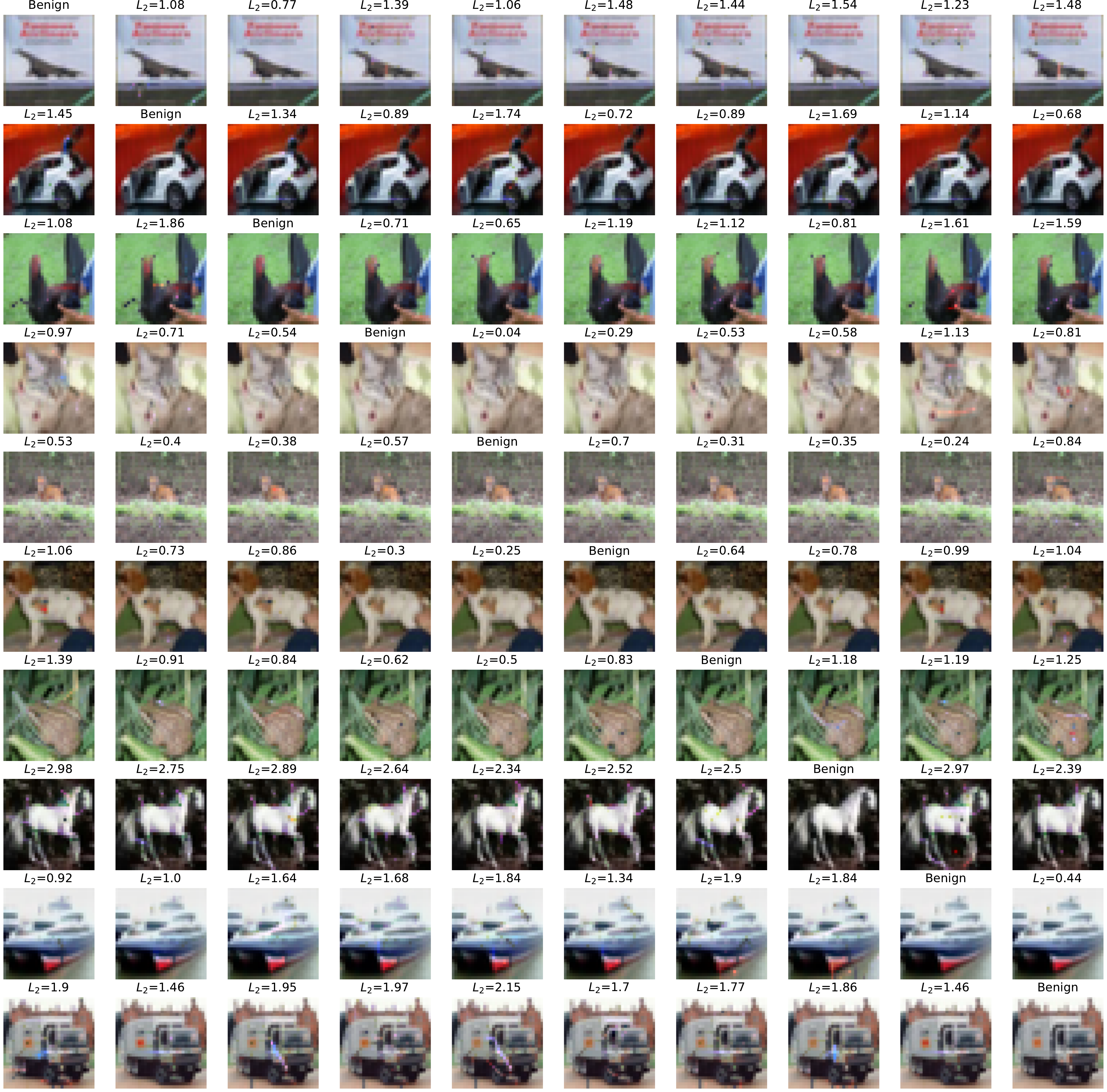}
    \caption{Targeted adversarial examples for each label pair of images on the adversarially trained model, $\kappa=0$}
    \label{fig:secret_k0}
\end{figure*}

\begin{figure*}
    \centering
    \includegraphics[width=0.95\textwidth]{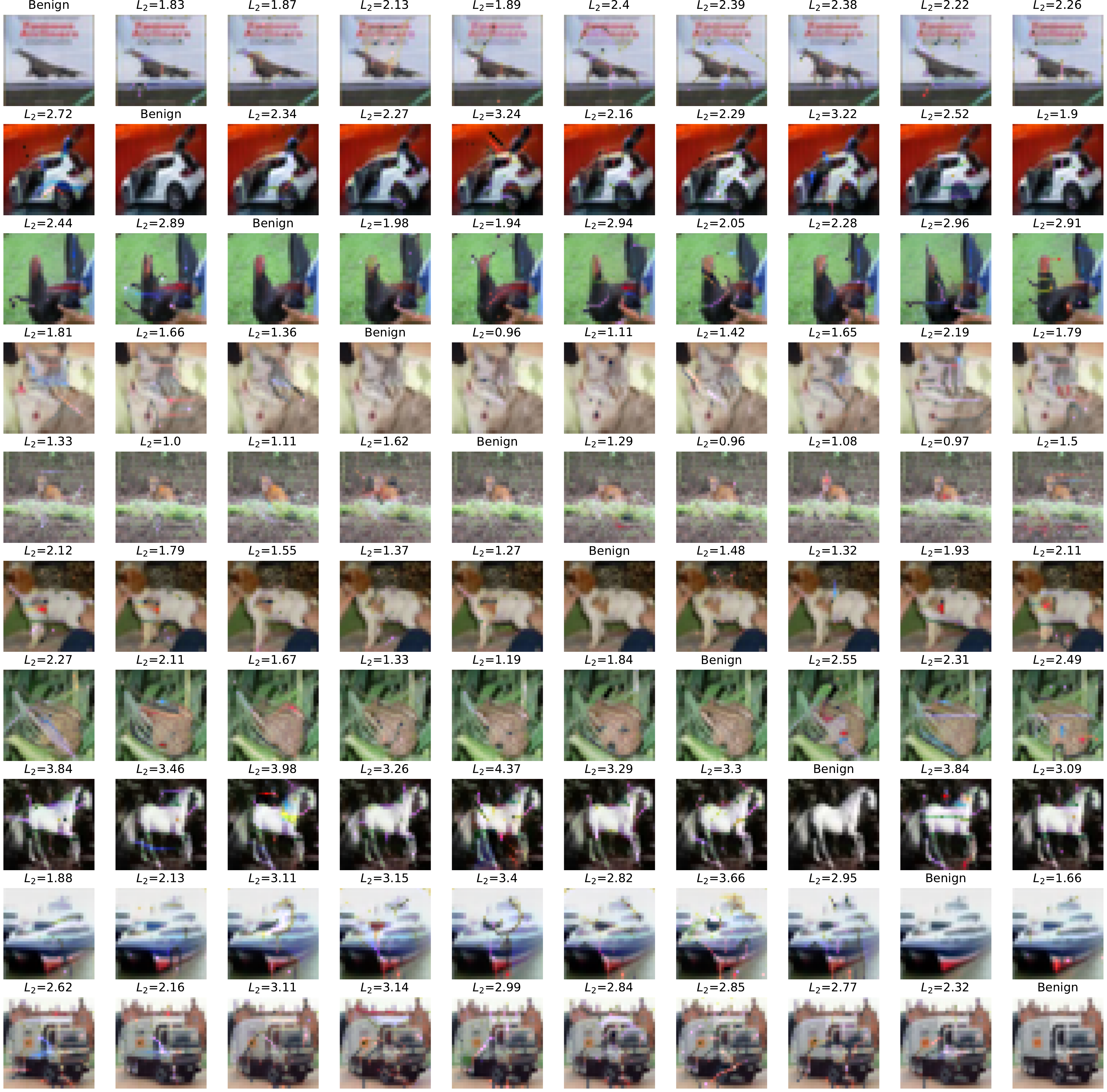}
    \caption{Targeted adversarial examples for each label pair of images on the adversarially trained model, $\kappa=10$}
    \label{fig:secret_k10}
\end{figure*}

Figures~\ref{fig:secret_k0} and \ref{fig:secret_k10} show adversarial examples on the adversarially training model with
$\kappa=0$ and $\kappa=10$ in Section~\ref{sec:a2d+at}, where each of them except for the benign ones targets
one of the other labels.
All the adversarial examples in Figure~\ref{fig:secret_k0} (i.e., $\kappa=0$)
can be detected by our defense \tool.
Only 3 adversarial examples in Figure~\ref{fig:secret_k10} (i.e., $\kappa=10$)
cannot be detected by \tool + AT.

%
%

\end{document}